\documentclass[twocolumn,superscriptaddress,amsmath,amssymb,pra]{revtex4-2}
\usepackage[utf8]{inputenc}
\usepackage[T1]{fontenc}
\usepackage{graphicx}
\usepackage{dcolumn}
\usepackage{appendix}
\usepackage{bm}
\usepackage{ulem}
\usepackage[usenames,dvipsnames]{xcolor}

\usepackage{subfigure}
\usepackage{bbm}
\usepackage{enumerate}

\usepackage[
bookmarks=true,
colorlinks,
linkcolor=black,
urlcolor=black,
citecolor=black,
plainpages=false,
pdfpagelabels,
final,
breaklinks=true
]{hyperref}

\usepackage{titlesec}

\usepackage{array}

\usepackage{physics}

\begin{document}
	\title{Propagation of intense squeezed vacuum light in non-linear media}
	
	\author{J. Rivera-Dean}
    \thanks{These authors contributed equally to this work}
	\affiliation{ICFO -- Institut de Ciencies Fotoniques, The Barcelona Institute of
		Science and Technology, Castelldefels (Barcelona) 08860, Spain}
	
	\author{D. Kanti}
    \thanks{These authors contributed equally to this work}
	\affiliation{Foundation for Research and Technology-Hellas, Institute of Electronic Structure \& Laser, GR-70013 Heraklion (Crete), Greece}
	
	\author{P. Stammer}
    \thanks{These authors contributed equally to this work}
	\affiliation{ICFO -- Institut de Ciencies Fotoniques, The Barcelona Institute of
		Science and Technology, Castelldefels (Barcelona) 08860, Spain}
	\affiliation{Atominstitut, Technische Universit\"{a}t Wien, 1020 Vienna, Austria}

    \author{S. Carlström}
    \affiliation{Max-Born-Institut für Nichtlineare Optik und Kurzzeitspektroskopie, im Forschungsverbund Berlin e.V. Max-Born-Straße 2 A, 12489 Berlin}
    
	\author{N. Tsatrafyllis}
	\affiliation{Foundation for Research and Technology-Hellas, Institute of Electronic Structure \& Laser, GR-70013 Heraklion (Crete), Greece}

    \author{M. Yu Ivanov}
    \affiliation{Max-Born-Institut für Nichtlineare Optik und Kurzzeitspektroskopie, im Forschungsverbund Berlin e.V. Max-Born-Straße 2 A, 12489 Berlin}
	
	\author{M. Lewenstein}
	\affiliation{ICFO -- Institut de Ciencies Fotoniques, The Barcelona Institute of
		Science and Technology, Castelldefels (Barcelona) 08860, Spain}
	\affiliation{ICREA, Pg. Lluís Companys 23, 08010 Barcelona, Spain}

	\author{P. Tzallas}
	\email{ptzallas@iesl.forth.gr}
	\affiliation{Foundation for Research and Technology-Hellas, Institute of Electronic Structure \& Laser, GR-70013 Heraklion (Crete), Greece}
	\affiliation{Center for Quantum Science and Technologies (FORTH-QuTech), GR-70013 Heraklion (Crete), Greece.}
	\affiliation{ELI-ALPS, ELI-Hu Non-Profit Ltd., Dugonics tér 13, H-6720 Szeged, Hungary}
	
	\date{\today}    
	
\begin{abstract}
     Recent developments in quantum light engineering have enabled the use of infrared bright squeezed vacuum (BSV) femtosecond pulses in highly nonlinear optics, particularly strong field physics and high-harmonic generation. However, theoretical studies were focused on the microscopic interaction with a single atom, neglecting the crucial macroscopic aspect of light propagation through the media. This raises a key question: How does BSV propagates in strongly light–driven nonlinear media and how this affects the generation of non-linear optical signals? We address this question by introducing a fully quantized framework that accounts for the propagation in gas media. We find that atomic ionization caused by strong BSV fluctuations and the associated infrared photon losses introduce decoherence effects that can substantially limit the propagation length in the medium, reduce the harmonic yield, and decrease the number of emitted harmonics at high intensities. However, these effects are not detrimental. We identify conditions under which propagation-induced decoherence is minimized while the generated harmonics remain clearly detectable--an issue of particular importance for future studies exploring the connection between strong-field physics and quantum optics. Our results lay the foundation for future studies of BSV in strong-field physics, nonlinear optics, and ultrafast science, and establish a basis for exploring its propagation through all states of matter in a fully quantized framework.
\end{abstract}

\date{\today}

\maketitle
\allowdisplaybreaks

Squeezed states of light, first developed decades ago, are one of the key resources for quantum technologies~\cite{Kimble_PRL_1986, Andersen_30Years_2016}. Recent advances in quantum light engineering have dramatically increased their brightness, enabling the generation of bright squeezed vacuum (BSV) states in the infrared (IR) regime~\cite{iskhakov2012polarization, Spasibko_PRL-BSV_2017, Manceau_PRL-BSV_2019}. Together with recent advantages in generating high-photon number non-classical states and multimode entangled light~\cite{lewenstein_generation_2021, rivera-dean_strong_2022, stammer_theory_2022, stammer_high_2022, Javier_PRA-ATI_2022, Javier_PRA-H2_2024, Foldi_PRA_2024, Javier_PRB_2024, Gonoskov_PRB_2024, Lamprou_PRL-Nonlinear_2025, Madsen_PRL_2025, Yi_PRX_2025, Pizzi_NatPhys-CorrAtoms_2023, Andrianov_PRA-DenseMedia_2024, Lange_PRA-ElectCorr_2024, Theidel_PRXQuantum_2024, Adrianov_PRA_2025, Imai_arXiv_2025}, these sources form key pillars of a rapidly emerging field often referred to as strong-field quantum optics. Born out of the synergy between quantum optics and strong laser-field physics, this field aims to merge tools and methods of these disciplines to both develop new approaches for fundamental research and novel applications in quantum information and ultrafast science at the fully quantized level ~\cite{Yi_PRX_2025, stammer_quantum_2023, Bhattacharya_2023, cruz-rodriguez_quantum_2024,Lamprou_JPB-Review_2025}.

BSV has already been employed in nonlinear optics for the generation of low-order harmonics in optical crystals~\cite{Spasibko_PRL-BSV_2017}, multiphoton~\cite{Heimerl_NatPhys-MultiElectrn_2024} and tunneling-induced~\cite{heimerl_driving_2025} electron emission, atomic ionization~\cite{Liu_PRL-AtomIon_2025, Lyu_PRR_2025}, and to probe the role of statistical properties of light in atomic spectroscopy~\cite{Mouloudakis_PRA-Spectr_2019}. High-photon number squeezed states, and in particular BSV states, have also emerged as promising drivers of extreme nonlinear processes, in particular high-harmonic generation (HHG)~\cite{Gorlach_NatPhys-HHG-BSV_2023, Wang_PRR-HHG-BSV_2024, Stammer_NatPhys-HHG-BSV_2024, Tzur_PRR-SqeezedHHG_2024, Javier_PRL-BSV_2025, Lemieux_NatPhoton-BSV_2025, Rasputnyi_NatPhys-BSV_2024,gothelf2025high, tzur_measuring_2025, stammer_weak_2025}, where theoretical studies have shown that IR squeezed light can produce harmonics with photon energies beyond those of conventional laser pulses~\cite{Wang_PRR-HHG-BSV_2024, Gorlach_NatPhys-HHG-BSV_2023, Javier_PRL-BSV_2025}. Moreover, it has been recently proposed that squeezed coherent states over a wide spectral range can be engineered via laser interactions inducing strong ground-state depletion~\cite{Stammer_PRL-Squeezing_2024}, through atoms resonantly excited or initially prepared in excited states~\cite{Yi_PRX_2025,rivera-dean_squeezed_2024}, or by driving HHG in strongly correlated materials~\cite{Lange_PRA-ElectCorr_2024,lange_hierarchy_2025,Madsen_PRL_2025}.

So far, however, theoretical investigations of BSV-driven HHG have focused on the single-atom response. ~As emphasized in Refs.~\cite{Tzallas_NatPhoton-NewsViews_2024, Lamprou_JPB-Review_2025}, assessing the applicability of such light sources requires incorporating the interplay of essential mechanisms such as ionization, ground state depletion, and field propagation through the medium.~While these aspects have been extensively studied for intense coherent femtosecond laser drivers using semiclassical approaches~\cite{Lewenstein_PRA-3step_1994, Salieres_AAMOP_1999, Constant_PRL_1999, Huillier_NatRevPhys_2022}, and for low photon number light sources using fully quantized approaches~\cite{Adam_PRA_1987, GlauberMaciej_PRA_1991, Jeffers_JMO_1994, Manzoni_NatCommun_2017, Raymer_JMO_2020}, they remain largely unexplored for intense BSV light sources due to the lack of fully quantized approaches capable of describing the propagation of intense BSV light in highly nonlinear media.

\begin{figure}
	\centering
	\includegraphics[width=0.9 \columnwidth]{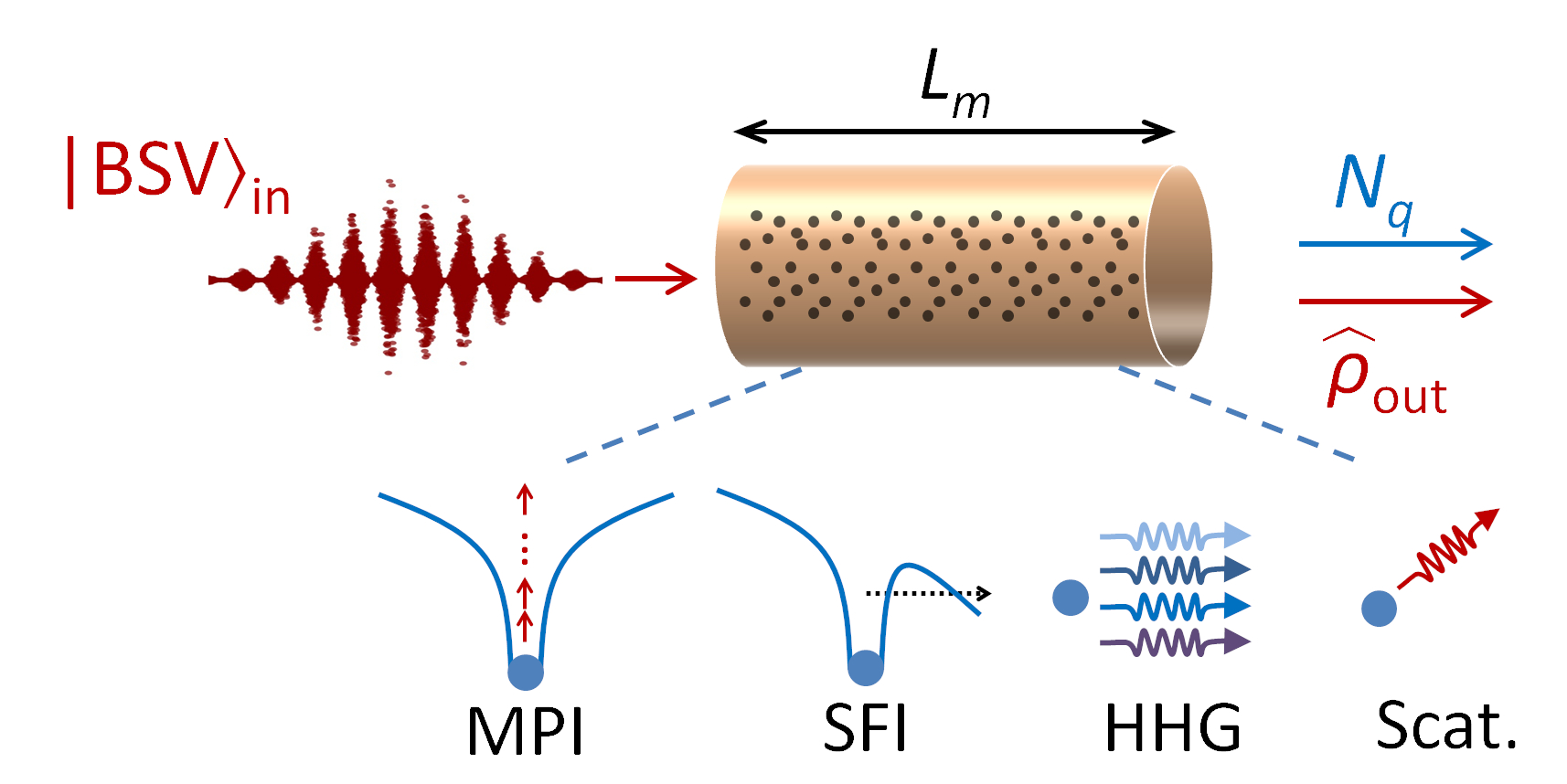}
	\caption{\textbf{Schematic of the propagation of intense IR BSV in a gas medium.} Multiphoton ionization (MPI), tunneling ionization (TI), high-harmonic generation (HHG), and IR scattering (Scat.) are the processes that typically occur during propagation.~$\ket{\text{BSV}}_{\text{in}}$ and $\hat {\rho}_{\text{out}}$ depict the IR light states entering and exiting a gas medium of length $L_{m}$, respectively. $N_{q}$ indicates the generated harmonic photon number.}
	\label{Fig.1}
\end{figure}

Here, we fill this gap  by introducing a fully quantized framework that accounts for the propagation of intense BSV in strongly driven nonlinear media. Focusing on a gas medium, we analyze conditions under which the BSV loses its quantum character due to fundamental processes occurring during propagation:   strong-field ionization (SFI), whether in multiphoton  (MPI) or tunneling (TI) regimes, scattering, and high harmonic generation (HHG) (Fig.~\ref{Fig.1}). We consider the interaction of Argon (Ar) atoms with intense, linearly polarized BSV light carried at $\lambda = 800$ nm, for multi-cycle femtosecond pulses.~Using HHG as a nonlinear observable, we reveal the central role of decoherence, the probabilistic nature of nonlinear interactions, the atomic ionization and ground-state depletion induced by the BSV pulse. Under conditions that preserve the quantum properties of BSV, we determine the number of high-harmonic photons generated and benchmark the results against those obtained with conventional coherent laser light sources.

\subsection*{Atomic ionization in a BSV light field}
Starting from single-atom interactions, one of the most fundamental steps in HHG is atomic ionization, whose rate we denote as $\Gamma$, and that generally depends on the time-dependent field intensity $\Gamma[I_\alpha(t)]$. With this, and for an arbitrary quantum state of light $\hat{\rho}$, the ionization probability reads
\begin{equation}\label{Eq:IonYield}
	\langle Y_{i}(t) \rangle=1-\int \dd E_\alpha Q(E_\alpha) 	e^{-\int_{-\infty}^{t}{\Gamma[I_\alpha(t)]dt}},
\end{equation}
where $Q(E_\alpha)$ corresponds to a marginal of the Husimi distribution, which represents the probability density of finding the quantum state $\hat{\rho}$ in the coherent state $\ket{\alpha}$. Equivalently, $Q(E_\alpha)$ gives the probability of having an electric field amplitude $E_\alpha$ (with $E_{\alpha} \propto \alpha$) and intensity $I_{\alpha}=\frac{c\epsilon_{0}}{2} E_{\alpha}^2$  (where $c$ is the speed of light in vacuum and $\epsilon_0$ is the vacuum permittivity) within $\hat{\rho}$. In our case, the rate has been calculated using time-dependent Schrödinger equation (TDSE) including all valence electrons in the $3s^2$ and $3p^6$-shells (see Appendix~\ref{Sec:App:Ion:rates}). Hereafter, for reasons of simplicity, we omit the subscript $\alpha$ from the amplitude distribution, i.e. $I_{\alpha}\equiv I$ and $E_{\alpha}\equiv E$.

\begin{figure}
	\centering
	\includegraphics[width=1.0 \linewidth]{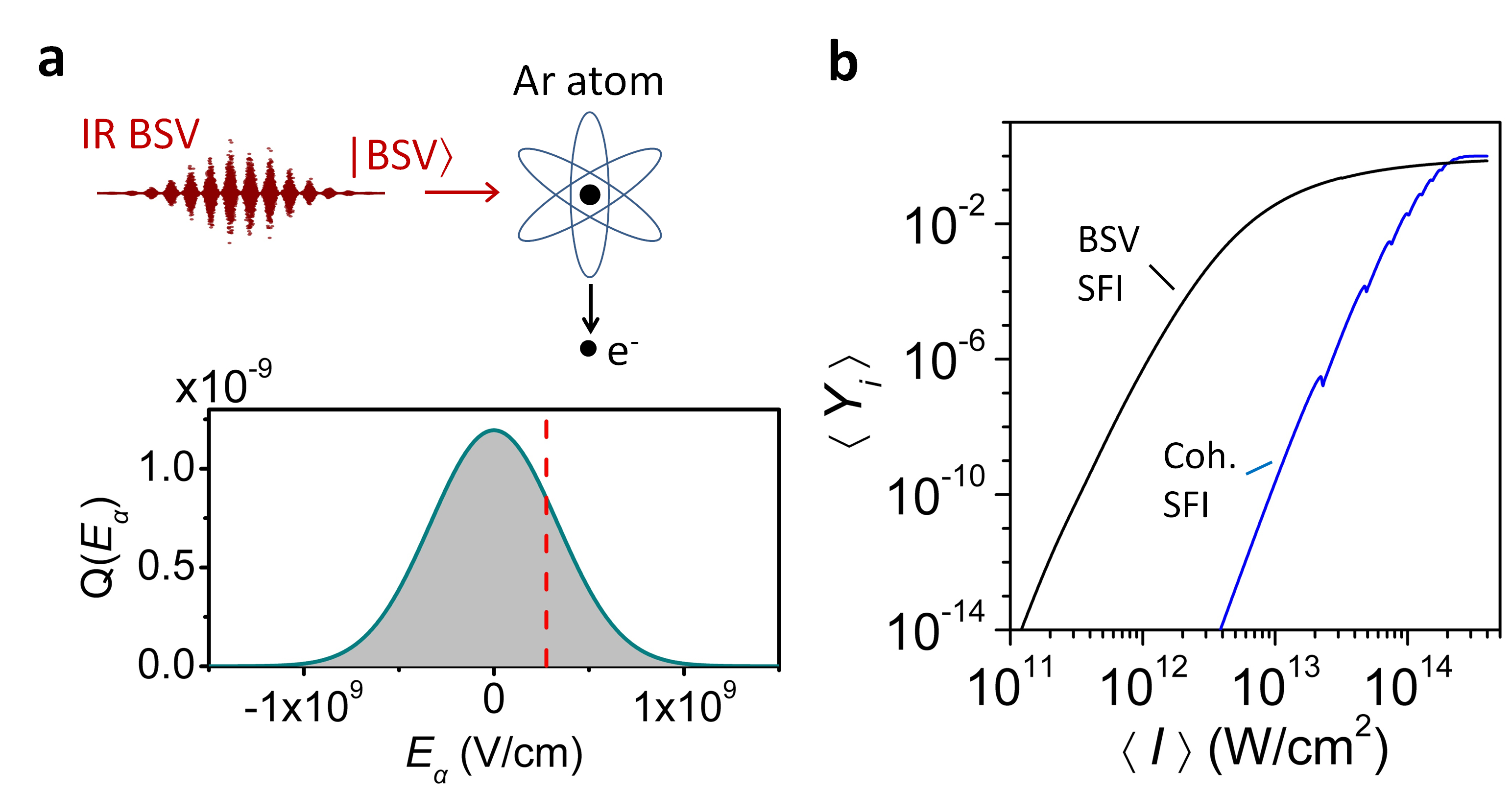}
	\caption{\textbf{Atomic ionization in a BSV field.} (\textbf{a}) Schematic of the interaction between BSV light pulses and a single Ar atom. $Q(E_{\alpha})$ represents a marginal of the Husimi distribution of BSV light with a mean intensity $\langle I\rangle = 1.3 \times 10^{14} \text{W}/\text{cm}^{2}$. For a coherent state of the same intensity $Q(E_{\alpha})$ is narrowly peaked at $E_{\alpha} \approx 3 \times 10^{8}$ (red-dashed line). (\textbf{b}) Dependence of the mean ionization probability $\langle Y_{i} \rangle$ on $\langle I\rangle $ of a BSV field (black line). For comparison, the dependence of $\langle Y_{i} \rangle$ on $\langle I\rangle $ for interactions with coherent light states is shown in blue.}
	\label{Fig.2}
\end{figure}

One of the defining features of $Q(E_{\alpha})$ for BSV light is its remarkably broad nature (Fig.~\ref{Fig.2}(a)) for $\langle I \rangle = 1.3 \times 10^{14}$ W/cm$^2$) which reflects the intrinsic field fluctuations of high-intensity squeezed light along the antisqueezed quadrature.~This behavior contrasts sharply with that of coherent states, whose distribution under the same mean intensity conditions is narrowly peaked at $E \approx 3 \times 10^{8}$ V/cm.~Such distinction has direct experimental implications:~while coherent light sources yield nearly identical intensities across successive laser shots, BSV pulses exhibit much larger shot-to-shot fluctuations, as encoded in $Q(E)$.~On average, these fluctuations follow $\langle I \rangle \propto \sinh^2(r)$, where $r$ is the squeezing amplitude, roughly corresponding to $r \approx 17$ for obtaining $\langle I\rangle \approx 1.3 \times 10^{14}$ W/cm$^2$ under experimental conditions.~This pronounced broadening of the field distribution is expected to strongly influence atomic ionization.

To illustrate this effect, Fig.~\ref{Fig.2}(b) shows the total ionization yield for both coherent (blue) and BSV (in black) light interacting with Ar atoms. The driving field is modeled as a pulse $E(t) = E f(t) \cos(\omega t + \theta)$, with $f(t)$ taken as a $\sin^2$-envelope of duration $\tau = 13$ fs. For interactions with coherent light states the ionization yield $\langle Y_{i} \rangle \approx 2\%$ occurs at intensities $\approx 1.3 \times 10^{14}$ W/cm$^2$, while for interactions with BSV, due to the enlarged field fluctuations, the ionization is substantially higher at much lower intensities providing the same ionization yield at intensities $\approx 9 \times 10^{12}$ W/cm$^2$. In the following sections, we discuss how this strong ionization induced by the BSV significantly affects both the quantum properties of the BSV field and the HHG process during light propagation in a medium.

\subsection*{Decoherence effects of the BSV field in non-linear media}
Quantum features of non-classical light fields are highly sensitive to decoherence effects arising from photon losses caused by scattering and absorption as light propagates through a medium. In the interaction of BSV with gaseous media, the majority of IR photon losses originate from absorption due to strong ionization, while those caused by scattering and the HHG process itself can be considered negligible (see Appendix~\ref{Sec:App:BSV:photon:losses}). The strong ionization modifies the quantum properties of the BSV state, transforming the initial pure state of the driving field $\ket{\text{BSV}}_{\text{in}}$ to a generally mixed state $\hat{\rho}_{\text{out}}$ (Fig.~\ref{Fig.1}). 

\begin{figure}
    \centering
    \includegraphics[width=1\columnwidth]{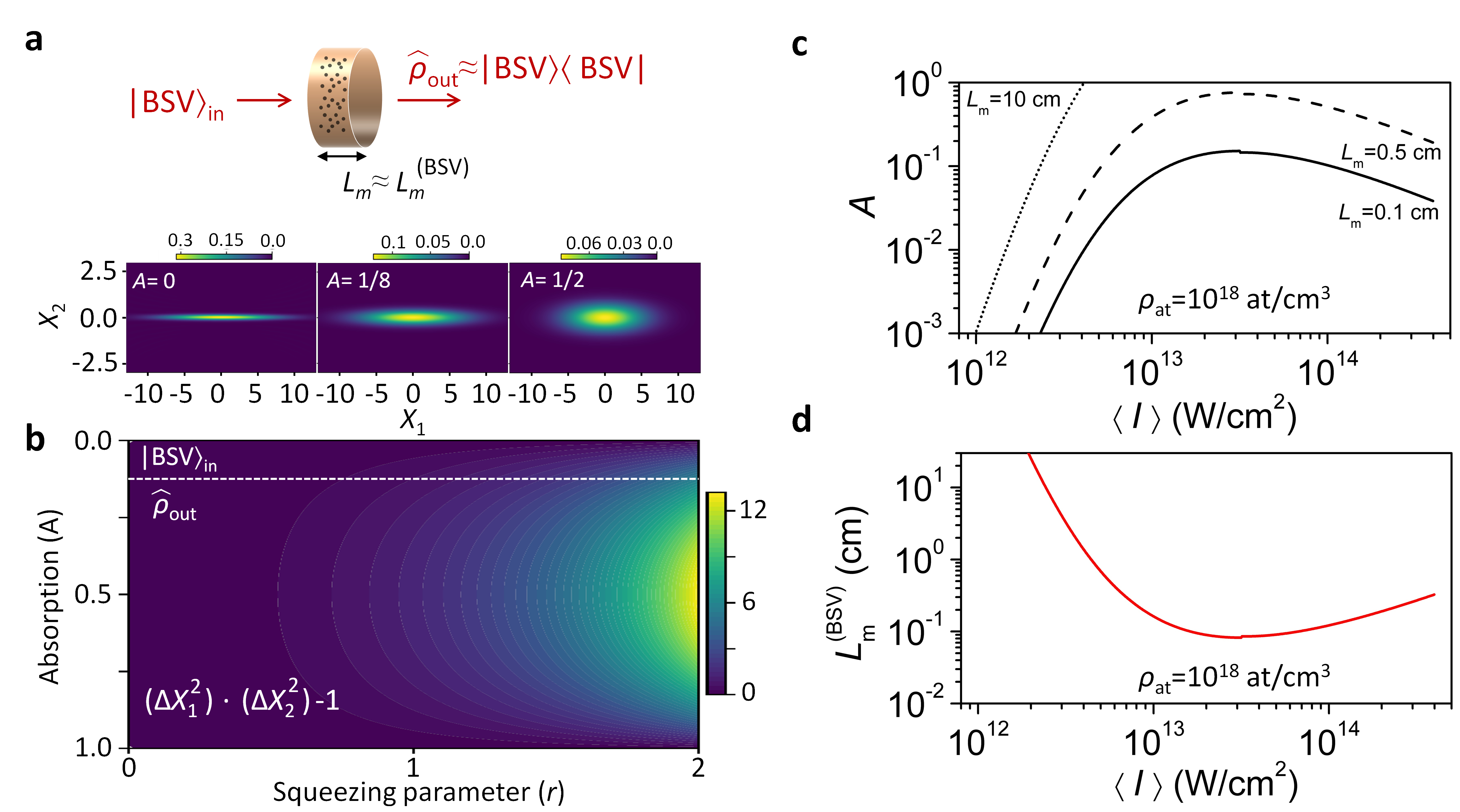}
    \caption{\textbf{Decoherence of BSV in non-linear media.} (\textbf{a}) Dependence of the Wigner function $W(X_{1},X_{2})$ of the light state on the photon losses during the propagation in the medium. Here, $\hat{X}_{1}=(\hat{a}^\dagger+\hat{a})$ and $\hat{X}_{2}=-i(\hat{a}^\dagger-\hat{a})$ are the optical quadrature operators. The left panel corresponds to the $W(X_{1},X_{2})$ of an initial BSV state $\ket{\text{BSV}}_{\text{in}}$, with $A=0$ ($0\%$ losses) and squeezing parameter $r$=2. The middle and right panels show the $W(X_{1},X_{2})$ for $A=1/8$ ($12.5\%$ photon losses) and $A=1/2$ ($50\%$ photon losses), respectively. (\textbf{b}) Dependence of product of variances of the optical quadratures on the photon losses minus the Heisenberg limit. The white dashed line depicts the $12\%$ photon losses that we define as the border above which the BSV state losses its quantumness, i.e., for $A<1/8$ the BSV preserves the initial quantum features $\ket{\text{BSV}}_{\text{in}}$ and for $A>1/8$ the state changes to a mixed state $\hat{\rho}_{\text{out}}$. (\textbf{c}) Dependence of $A$ on $\langle I \rangle$ at different values of $L_{m}$ for $\rho_{\text{at}}=10^{18}$ atoms per cm$^3$. (\textbf{d}) Dependence of $L_{m}^{\text{(BSV)}}$ on $\langle I \rangle$ for $\rho_{\text{at}}=10^{18}$ atoms per cm$^3$.}
    \label{Fig.3}
\end{figure}

To quantify this impact, we analyze the evolution of the Wigner function $W(X_1,X_2)$ (Fig.~\ref{Fig.3}(a)) and the deviation of the product of the optical quadrature variances from the Heisenberg uncertainty limit, i.e., $(\Delta X_{1}^2)(\Delta X_{2}^2)-1$ (Fig.~\ref{Fig.3}(b)) under photon losses~\cite{leonhardt_quantum_1993} (see Appendix~\ref{Sec:App:g2:and:quadratures}). The optical quadratures are defined as $\hat{X}_1 = (\hat{a}^\dagger + \hat{a})$ and $\hat{X}_2 = i(\hat{a}-\hat{a}^\dagger)$. At extreme absorption ($A$) values, the product reaches the Heisenberg limit, with $\hat{\rho}_{\text{out}}$ corresponding either to a pure BSV (no absorption $A=0$) or a vacuum state (maximum absorption, $A=1$).~For intermediate losses, the state is noisy: although the quadrature distributions retain some stretching reminiscent of squeezing, these features do not correspond to genuine quantum squeezing below the vacuum level.~Based on this, we define that the BSV loses its quantumness when photon losses exceed $A=1/8$ ($12.5\%$ of the mean IR photon number $\langle N_{\text{IR}} \rangle$) (white dashed line in Fig.~\ref{Fig.3}(b)).~Considering that $n$ IR photons are required to ionize a single atom, the number of absorbed IR BSV photons as a function of the medium length $L_{m}$ can be roughly estimated as $\langle N_{\text{abs}} \rangle= \langle Y_{i} \rangle n  \rho_{\text{at}}S_{\text{BSV}}L_{m}$. $S_{\text{BSV}}$ is the focal spot area of a BSV light beam of $\langle I \rangle  = \hbar\omega \langle N_{\text{IR}} \rangle/\tau S_{\text{BSV}}$ and $\rho_{\text{at}}$ is the atomic density. Based on this, we can estimate the absorption factor $A=\langle N_{\text{abs}} \rangle/\langle N_{\text{IR}} \rangle$ due to strong-field ionization as
\begin{equation}\label{Eq:Photonlosses}
	\begin{split}
		A= \frac{\langle Y_{i} \rangle n  \rho_{\text{at}}S_{\text{BSV}}L_{m}}{\langle N_{\text{IR}}\rangle}=\frac{\hbar \omega\langle Y_{i} \rangle n  \rho_{\text{at}}L_{m}}{\tau \langle I\rangle},
	\end{split}
\end{equation} 
with the condition that $A \leq 1$. Figure~\ref{Fig.3}(c) illustrates the dependence of $A$ on $\langle I \rangle$ at different values of $L_{m}$ for $\rho_{\text{at}}=10^{18}$ atoms per cm$^3$. Using Eq.~\ref{Eq:Photonlosses} and considering the limit $A\lesssim 1/8$, a relation which provides an upper medium length ($L_{m}^{\text{(BSV)}}$) upon which the BSV preserves its \textit{quantumness} is,  
\begin{equation}\label{Eq:PropLength}
	\begin{split}
		L_{m}\lesssim \frac{\tau \langle I \rangle}{8\hbar\omega\langle Y_{i} \rangle n\rho_{\text{at}}}=L_{m}^{(\text{BSV})}.
	\end{split}
\end{equation}
The dependence of $L_{m}^{\text{(BSV)}}$ on $\langle I \rangle$ for $\rho_{\text{at}}=10^{18}$ atoms per cm$^3$ is shown in Fig.~\ref{Fig.3}(d).

\begin{figure*}
    \centering
    \includegraphics[width=1\textwidth]{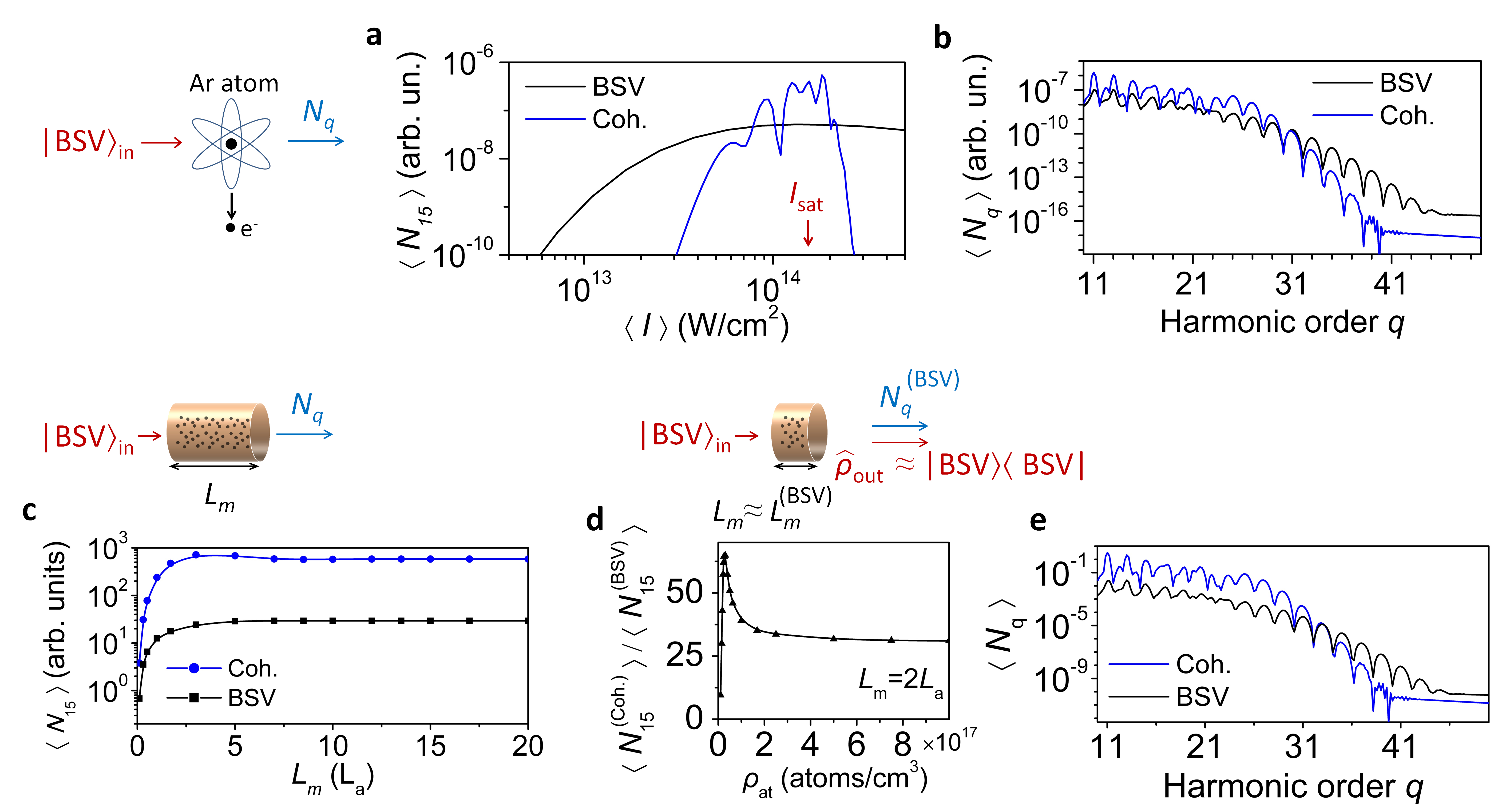}
    \caption{\textbf{Propagation of BSV in non-linear media and HHG.} The upper panels ((\textbf{a}), (\textbf{b})) refer to the harmonic yield generated when a BSV and coherent fields interact with a single Ar atoms. The lower panels ((\textbf{c}), (\textbf{b}) and (\textbf{e})) refer to the harmonic yield exiting an Ar gas medium when the interaction occurs with BSV and coherent fields.   (\textbf{a}) Dependence of the mean photon number of the 15th harmonic $\langle N_{15} \rangle$ on $\langle I\rangle $ of a BSV field (black line). For comparison, the dependence of $\langle N_{15} \rangle$ on $\langle I\rangle $ for interactions with coherent light states is shown with a blue line. (\textbf{b}) HHG spectra generated by BSV (black line) and coherent (blue line) fields at $\langle I\rangle =  1.3 \times 10^{14} \text{W}/\text{cm}^{2}$. (\textbf{c}) Dependence of $\langle N_{15} \rangle$ on the medium length $L_{m}$ generated by BSV (black line) and coherent (blue line) fields without taking into account the BSV decoherence effects in the medium. (\textbf{d}) Dependence of  $\langle N_{15} \rangle^{(\text{Coh.})}/\langle N_{15} \rangle^{(\text{BSV})}$ on $\rho_{\text{at}}$ calculated taking into account the BSV decoherence in the medium. The calculation has been conducted for $L_{m}=2L_{a}$ where the BSV preserves its quantumness. For the coherent states the calculation has been conducted for $L_{m}>\frac{5}{2}  L_{a}$.  (\textbf{e}) HHG spectra generated by BSV (black line) and coherent states (blue line) after propagation of $L_{m}\approx L_{m}^{\text{(BSV)}}\approx 2 L_{a}$ and $L_{m}>\frac{5}{2}  L_{a}$, respectively. $N_{q}^{\text{(BSV)}}$ indicates the harmonic photon number generated by the state $\hat{\rho}_{\text{out}}\approx \dyad{\text{BSV}}$.}
    \label{Fig.4}
\end{figure*}

\subsection*{Propagation of BSV in non-linear media and HHG}
Having established the interplay between BSV driving fields and strong-field ionization, we now turn to the HHG response. For single-atom interactions, Fig.~\ref{Fig.4}(a) shows the harmonic yield $\langle N_q\rangle$ of the 15th harmonic order as a function of $\langle I \rangle$ for both types of fields. In the case of BSV-driven interactions, $\langle N_q\rangle$ reaches its maximum within the intensity range $\approx 6 \times 10^{13}$ to $\approx 2 \times 10^{14}$ W/cm$^2$. For coherent light, the corresponding range is narrower, exhibiting a peak around $\approx 1.3\times 10^{14}$ W/cm$^2$. Hereafter, we define $I_{\text{sat}}=1.3\times 10^{14}$ W/cm$^2$ as the harmonic saturation intensity, i.e., the value yielding the maximum harmonic output for both fields. This intensity range is commonly employed in HHG and attosecond experiments using the interaction of Ar atoms with intense coherent light~\cite{Constant_PRL_1999,Hergott_PRA_2002, Huillier_NatRevPhys_2022}, as it provides maximum harmonic yield and while keeps the ionization at low levels $\langle Y_{i} \rangle \approx 2\%$ (Fig.~\ref{Fig.2}(b)), conditions which is important for maintaining the phase matching conditions and attosecond pulse formation. For a BSV field at this intensity range, although the harmonic yield is maximum, the ionization level is also high $\langle Y_{i} \rangle \approx 50\%$ (Fig.~\ref{Fig.2}(b)), an issue that substantially affects the quantum properties of the BSV and the harmonic yield exiting the medium. We also, restrict our analysis to this regime, since we can roughly consider that for both fields (BSV and coherent), at higher intensities the yield decreases due to ground-state depletion, while at lower intensities is suppressed by the HHG nonlinearity.

In this analysis, the mean photon number of the $q$th harmonic $\langle N_{q} \rangle$ is evaluated as $\langle N_q\rangle = \int \dd E Q(E) \lvert\text{FT}[\langle \psi_E(t) \vert \hat{d}\vert \psi_E(t)\rangle]\rvert^2$, where $\hat{d}$ is the dipole operator, FT[$\cdot$] denotes the Fourier Transform, and $\ket{\psi_E(t)}$ is the electron state driven by $E(t)$, which incorporates ground-state depletion via the ionization rate $\Gamma$ (see Appendix~\ref{Sec:App:HHG:Spec}). For BSV drivers, this approach effectively neglects quantum coherence between the different coherent-state realizations of the BSV, especially natural in the case of strong absorption of the driving
field by the medium. This absorption inevitably entangles light with medium during propagation. Thus, as long as the observation traces out the final state of the generating medium, the quantum coherence between the different BSV realizations will be lost.

At low intensities $\langle I \rangle \ll I_{\text{sat}}$ the BSV-driven harmonics dominate,  thanks to the enhanced field fluctuations which allow the BSV to sample higher intensities. At $\langle I \rangle \sim I_{\text{sat}}$ the single-atom harmonic yield is already strongly  suppressed by the strong ground-state depletion, again due to the high intensity fluctuations of the BSV. As a result, around $I_{\text{sat}}$ the single-atom HHG yield for the BSV driver is over an order of magnitude lower than for the  coherent state driver, with comparable values observed across other plateau harmonics (not shown in Fig.~\ref{Fig.4}(a) for simplicity). Once $\langle I \rangle > I_{\text{sat}}$, depletion effects dominate, leading to a rapid reduction of the HHG field. The suppression is much faster for the coherent state driver, as now the BSV-driven atoms benefit from sampling lower intensities. Notably, while HHG driven by coherent states exhibits an oscillatory behavior in the yield due to quantum path interference~\cite{Zair_PRA_2008}, these interferences are washed out by the  intensity fluctuations of the BSV driver.

This broadening of the electric field amplitudes also explains the, relatively smaller here due to the ground state depletion, extension of the highest harmonics observed for BSV compared to coherent sources~\cite{Gorlach_NatPhys-HHG-BSV_2023}.~Figure~\ref{Fig.4}(b) shows the HHG spectrum for both fields at $\langle I \rangle  = I_{\text{sat}}$, with a cutoff at the 29th harmonic for coherent light and at the 33rd for BSV. We note, however, that these spectra would correspond to multi-shot averages over single-atom responses and, at this stage, do not include propagation or decoherence effects. Importantly, while $N_{q}(I)$ and $Y_{i}(I)$ at a specific intensity $I$ cannot be observed experimentally for a BSV  driver, they play crucial role in the propagation effects and the resulting decoherence. These effects are unavoidable in any experimental configuration and significantly influence the emitted HHG radiation.

To account for propagation of the harmonic emission to the detector, we take advantage of the fact that,  under strong BSV driving, the microscopic state of the $q$th harmonic can be approximated by the mixed state $\hat{\rho}_q = \int \dd E Q(E) \dyad{\chi_q(E)}$, where $\chi_{q}(E)$ represents a coherent state of amplitude $\chi_q$ when the process is driven by the field $E(t)$~\cite{Gorlach_NatPhys-HHG-BSV_2023,Javier_PRL-BSV_2025}. Then, we incorporate on-axis propagation of these states in our calculations, and find that the intensity of the $q$th harmonic mode at time $t$ and position $z$ after propagation reads
\begin{equation} \label{Eq:Maxwell}
    \begin{aligned}
        \langle I_q\rangle 
            &= \int \dd E\  Q(E) \tr[\hat{E}^2_q(z,t)\hat{\rho}(t)]
            \\& \approx \int \dd E\  Q(E) E^{2}_q(z,t;\alpha), 
    \end{aligned}
\end{equation}
with $E_q(z,t)$ obeying the propagation equation under the paraxial approximation (see Appendix~\ref{Sec:App:HHG:Spec}).~This allows to calculate the generated harmonic photon number by introducing, via $Q(E)$, the amplitude distribution of the BSV driving field the into the standard semi-classical propagation equations~\cite{Constant_PRL_1999}.~For simplicity, we assume that the microscopic atomic response amplitude is independent of the position in the medium---an assumption which is valid when the IR confocal parameter ($d_{f}$) greatly exceeds the medium length $L_{m}$. We also set the mean intensity to $\langle I\rangle =I_{\text{sat}}$.

Under these conditions, the mean photon number of the $q$th harmonic mode exiting on-axis from a gas medium with atomic density $\varrho_{\text{at}}$ reads
\begin{equation}\label{Eq:Propagation}
	\begin{split}
		\langle N_{q} \rangle \propto  \bigg\langle B N_{q}( I ) \left[1+e^{-L'}-2\cos(\frac{\pi L_{m}}{L_{c}}) e^{-\frac{L'}{2}} \right] \bigg\rangle,
	\end{split}
\end{equation}
where $B=4\varrho_{\text{at}}^{2} L_{a}/(1+4\pi^{2}(L_{a}/L_{c})^{2})$ and $L'=L_{m}/L_{a}$. Here, $L_{a}=1/(\sigma^{(1)}\varrho_{\text{at}})$ is the absorption length, while $L_{c}=\pi/|\Delta k|$ is the coherence length, and $\Delta k$ is the phase mismatch (see Appendix~\ref{Sec:App:HHG:Spec}). Finally, $N_{q}$ depends on the quantum properties of the driving field via $N_{q}(I)$ and $L_{c} (I) \propto f(Y_{i}(I))$, both associated to $Q(E)$. We illustrate the calculations using the 15th harmonic, and a gas density of $\rho_{\text{at}} \sim 10^{18}$ atoms/cm$^3$, typical for HHG experiments(see Appendix~\ref{Sec:App:HHG:Spec}). 

Finally, Fig.~\ref{Fig.4}(c) shows $\langle N_{15} \rangle$ as a function of the medium length $L_{m}$ for BSV (black-line) and coherent (blue--line) state driving fields. Two key features emerge: (i) for both BSV and coherent fields, $\langle N_{15} \rangle$ is maximized when $L_{m}>\frac{5}{2}  L_{a}$; (ii) in this regime ($L_{m}> \frac{5}{2} L_{a}$), the photon yield for coherent states exceeds that of BSV by about 30 times, i.e., $\langle N_{15} \rangle^{(\text{Coh.})}/\langle N_{15} \rangle^{(\text{BSV})}\sim 30$. Taking into account the decoherence effects and the limitations introduced by Eq.~\ref{Eq:PropLength}, which provide a maximum medium length of $L_m \approx L_{m}^{(\text{BSV})} \approx 2 L_{a}$, we obtain (Methods) that
\begin{equation}\label{Eq:HarmYield}
	\begin{split}
		\langle N_{q} \rangle^{(\text{Coh.})}/\langle N_{q} \rangle^{(\text{BSV})}\gtrsim  50,
	\end{split}
\end{equation}
indicating that, for $\rho_{\text{at}}>2\times10^{16}$ atoms per cm$^3$ (Fig.~\ref{Fig.4}(d)), the conventional coherent laser sources are by more than 50 times more efficient for the generation of plateau harmonics that the BSV states in the $\langle I \rangle = I_{\text{sat}}$ regime. In this case the corresponding harmonic spectra are shown in Fig.~\ref{Fig.4}(e) (Methods).

\subsection*{Discussion}
In recent theoretical studies involving the ideal case of interaction with a single atom, it has been shown that the harmonic yield and the number of generated harmonic frequencies increase dramatically when using BSV pulses compared to intense coherent light states \cite{Gorlach_NatPhys-HHG-BSV_2023}. However, our findings show that propagation effects in the medium, limit the practical use of intense BSV sources in strong-field physics and HHG compared to intense coherent light. These limitations arise from decoherence due to photon losses during propagation---primarily from enhanced ionization in the medium---as well as from the broader intensity distribution of BSV states. Consequently, the effective propagation length in the medium is reduced, and the probability of observing events with sufficiently high intensities to generate high harmonics is lower than for coherent states. Moreover, the nonlinearity of the processes driven by intense BSV light, combined with atomic ground-state depletion, further constraints the HHG yield. As a result of the above, when the HHG process is driven by BSV light in a gas medium, the number of the emitted harmonic frequencies is very similar to those produced by coherent light, while the harmonic yield is more than 50 times lower than that of the coherent light states.     

Nevertheless, the ability to detect a highly nonlinear observable, such as high harmonics, remains of significant interest for applications in ultrafast science and quantum technologies \cite{Bhattacharya_2023, cruz-rodriguez_quantum_2024, Lamprou_JPB-Review_2025}.~Thus, it is crucial to address the conditions at which the BSV harmonics are detectable. To quantify this, we used experimentally and theoretically established HHG conversion efficiencies for Ar atoms (see Appendix~\ref{Sec:App:HHG:Spec}). For plateau harmonics driven by a BSV driving field of mean photon number $\sim 10^{13}$ photons per pulse~\cite{Manceau_PRL-BSV_2019}, a medium length of $L_{m}^{(\text{BSV})}$, yields $\sim 5 \times 10^{2}$ harmonic photons per pulse exiting the Ar medium. Such photon numbers are detectable. 

An advantage of BSV light over standard coherent states emerges in the intensity range $\langle I \rangle < I_{\text{sat}}$---a regime in which coherent light states cannot produce harmonics. In this way, we can minimize or even eliminate decoherence effects, since photon losses caused by atomic ionization are significantly reduced (Figs.~\ref{Fig.3}(c), (d)), i.e., $A\ll 12\%$. For example, it can be estimated (see Appendix~\ref{Sec:App:BSV:photon:losses}) that for BSV light with $\langle I \rangle \sim 10^{13}$ W/cm$^2$, the emitted photon number is in the range of 20 photons per pulse, which remains detectable. Therefore, this intensity regime can also be considered as suitable for the generation of high harmonics without BSV light being affected by decoherence effects. It is furthermore important to emphasize that none of the results presented here can be taken as a proof of the quantum nature of a high-photon-number BSV source, or as evidence that quantum features are imprinted on the HHG observables. Demonstrating this would require methods capable of characterizing quantum states at high photon numbers while minimizing decoherence from passive optical elements—an issue beyond the scope of this work.

Since ionization, harmonic generation, and light scattering are ubiquitous in light–matter interactions, our study provides a foundation for understanding the propagation of intense BSV light across all states of matter, such us the recent works conducted in solids \cite{Rasputnyi_NatPhys-BSV_2024, Lemieux_NatPhoton-BSV_2025}. This underlines its relevance for future investigations in strong-field physics, nonlinear optics, and ultrafast science~\cite{cruz-rodriguez_quantum_2024,Lamprou_JPB-Review_2025}. Finally, these findings demonstrate the need to account for propagation effects when examining the phase locking of harmonics generated by BSV states, a consideration that is particularly critical for attosecond science applications.

\section*{Acknowledgements}

The group of P.T. at FORTH acknowledges the Hellenic Foundation for Research and Innovation (HFRI) and the General Secretariat for Research and Technology (GSRT) under grant agreement CO2toO2 Nr.:015922, the European Union’s HORIZON-MSCA-2023-DN-01 project QU-ATTO under the Marie Skłodowska-Curie grant agreement No 101168628 and ELI--ALPS. ELI--ALPS is supported by the EU and co-financed by the European Regional Development Fund (GINOP No. 2.3.6-15-2015-00001). H2020-EU research and innovation program under the Marie Skłodowska-Curie (No. 847517). ICFO-QOT group acknowledges support from: European Research Council AdG NOQIA; MCIN/AEI (PGC2018-0910.13039/501100011033,  CEX2019-000910-S/10.13039/501100011033, Plan National STAMEENA PID2022-139099NB, project funded by MCIN/AEI/10.13039/501100011033 and by the “European Union NextGenerationEU/PRTR" (PRTR-C17.I1), FPI); project funded by the EU Horizon 2020 FET-OPEN OPTOlogic, Grant No 899794, QU-ATTO, 101168628), Fundació Cellex; Fundació Mir-Puig; Generalitat de Catalunya (European Social Fund FEDER and CERCA program.
P.S. acknowledges funding from the European Union’s Horizon 2020 research and innovation program under the Marie Skłodowska-Curie grant agreement No 847517.

\bibliography{References.bib}{}

\begin{thebibliography}{73}%
\makeatletter
\providecommand \@ifxundefined [1]{%
 \@ifx{#1\undefined}
}%
\providecommand \@ifnum [1]{%
 \ifnum #1\expandafter \@firstoftwo
 \else \expandafter \@secondoftwo
 \fi
}%
\providecommand \@ifx [1]{%
 \ifx #1\expandafter \@firstoftwo
 \else \expandafter \@secondoftwo
 \fi
}%
\providecommand \natexlab [1]{#1}%
\providecommand \enquote  [1]{``#1''}%
\providecommand \bibnamefont  [1]{#1}%
\providecommand \bibfnamefont [1]{#1}%
\providecommand \citenamefont [1]{#1}%
\providecommand \href@noop [0]{\@secondoftwo}%
\providecommand \href [0]{\begingroup \@sanitize@url \@href}%
\providecommand \@href[1]{\@@startlink{#1}\@@href}%
\providecommand \@@href[1]{\endgroup#1\@@endlink}%
\providecommand \@sanitize@url [0]{\catcode `\\12\catcode `\$12\catcode
  `\&12\catcode `\#12\catcode `\^12\catcode `\_12\catcode `\%12\relax}%
\providecommand \@@startlink[1]{}%
\providecommand \@@endlink[0]{}%
\providecommand \url  [0]{\begingroup\@sanitize@url \@url }%
\providecommand \@url [1]{\endgroup\@href {#1}{\urlprefix }}%
\providecommand \urlprefix  [0]{URL }%
\providecommand \Eprint [0]{\href }%
\providecommand \doibase [0]{https://doi.org/}%
\providecommand \selectlanguage [0]{\@gobble}%
\providecommand \bibinfo  [0]{\@secondoftwo}%
\providecommand \bibfield  [0]{\@secondoftwo}%
\providecommand \translation [1]{[#1]}%
\providecommand \BibitemOpen [0]{}%
\providecommand \bibitemStop [0]{}%
\providecommand \bibitemNoStop [0]{.\EOS\space}%
\providecommand \EOS [0]{\spacefactor3000\relax}%
\providecommand \BibitemShut  [1]{\csname bibitem#1\endcsname}%
\let\auto@bib@innerbib\@empty
\bibitem [{\citenamefont {\relax{Ling-An Wu, H. J. Kimble, J. L. Hall, and
  Huifa Wu}}(1986)}]{Kimble_PRL_1986}%
  \BibitemOpen
  \bibfield  {author} {\bibinfo {author} {\bibnamefont {\relax{Ling-An Wu, H.
  J. Kimble, J. L. Hall, and Huifa Wu}}},\ }\bibfield  {title} {\bibinfo
  {title} {Generation of squeezed states by parametric down conversion},\
  }\href {https://doi.org/doi.org/10.1103/PhysRevLett.57.2520} {\bibfield
  {journal} {\bibinfo  {journal} {Phys. Rev. Lett.}\ }\textbf {\bibinfo
  {volume} {57}},\ \bibinfo {pages} {2520} (\bibinfo {year}
  {1986})}\BibitemShut {NoStop}%
\bibitem [{\citenamefont {\relax{U. L. Andersen, T. Gehring, C. Marquardt, and
  G. Leuchs}}(2016)}]{Andersen_30Years_2016}%
  \BibitemOpen
  \bibfield  {author} {\bibinfo {author} {\bibnamefont {\relax{U. L. Andersen,
  T. Gehring, C. Marquardt, and G. Leuchs}}},\ }\bibfield  {title} {\bibinfo
  {title} {30 years of squeezed light generation},\ }\href
  {https://doi.org/10.1088/0031-8949/91/5/053001} {\bibfield  {journal}
  {\bibinfo  {journal} {Phys. Scr.}\ }\textbf {\bibinfo {volume} {91}},\
  \bibinfo {pages} {053001} (\bibinfo {year} {2016})}\BibitemShut {NoStop}%
\bibitem [{\citenamefont {Iskhakov}\ \emph {et~al.}(2012)\citenamefont
  {Iskhakov}, \citenamefont {Agafonov}, \citenamefont {Chekhova},\ and\
  \citenamefont {Leuchs}}]{iskhakov2012polarization}%
  \BibitemOpen
  \bibfield  {author} {\bibinfo {author} {\bibfnamefont {T.~S.}\ \bibnamefont
  {Iskhakov}}, \bibinfo {author} {\bibfnamefont {I.~N.}\ \bibnamefont
  {Agafonov}}, \bibinfo {author} {\bibfnamefont {M.~V.}\ \bibnamefont
  {Chekhova}},\ and\ \bibinfo {author} {\bibfnamefont {G.}~\bibnamefont
  {Leuchs}},\ }\bibfield  {title} {\bibinfo {title} {Polarization-entangled
  light pulses of 10 5 photons},\ }\href@noop {} {\bibfield  {journal}
  {\bibinfo  {journal} {Physical Review Letters}\ }\textbf {\bibinfo {volume}
  {109}},\ \bibinfo {pages} {150502} (\bibinfo {year} {2012})}\BibitemShut
  {NoStop}%
\bibitem [{\citenamefont {{\relax{K. Y. Spasibko, D. A. Kopylov, V. L.
  Krutyanskiy, T. V. Murzina, G. Leuchs, and M. V.
  Chekhova}}}(2017)}]{Spasibko_PRL-BSV_2017}%
  \BibitemOpen
  \bibfield  {author} {\bibinfo {author} {\bibnamefont {{\relax{K. Y. Spasibko,
  D. A. Kopylov, V. L. Krutyanskiy, T. V. Murzina, G. Leuchs, and M. V.
  Chekhova}}}},\ }\bibfield  {title} {\bibinfo {title} {\relax{Multiphoton
  effects enhanced due to ultrafast photon-number fluctuations}},\ }\href
  {https://doi.org/...} {\bibfield  {journal} {\bibinfo  {journal} {Phys. Rev.
  Lett.}\ }\textbf {\bibinfo {volume} {119}},\ \bibinfo {pages} {223603}
  (\bibinfo {year} {2017})}\BibitemShut {NoStop}%
\bibitem [{\citenamefont {{\relax{M. Manceau, K. Y. Spasibko, G. Leuchs, R.
  Filip, and M. V. Chekhova}}}(2019)}]{Manceau_PRL-BSV_2019}%
  \BibitemOpen
  \bibfield  {author} {\bibinfo {author} {\bibnamefont {{\relax{M. Manceau, K.
  Y. Spasibko, G. Leuchs, R. Filip, and M. V. Chekhova}}}},\ }\bibfield
  {title} {\bibinfo {title} {\relax{Indefinite-mean pareto photon distribution
  from amplified quantum noise}},\ }\href {https://doi.org/...} {\bibfield
  {journal} {\bibinfo  {journal} {Phys. Rev. Lett.}\ }\textbf {\bibinfo
  {volume} {123}},\ \bibinfo {pages} {123606} (\bibinfo {year}
  {2019})}\BibitemShut {NoStop}%
\bibitem [{\citenamefont {Lewenstein}\ \emph {et~al.}(2021)\citenamefont
  {Lewenstein}, \citenamefont {Ciappina}, \citenamefont {Pisanty},
  \citenamefont {Rivera-Dean}, \citenamefont {Stammer}, \citenamefont
  {Lamprou},\ and\ \citenamefont {Tzallas}}]{lewenstein_generation_2021}%
  \BibitemOpen
  \bibfield  {author} {\bibinfo {author} {\bibfnamefont {M.}~\bibnamefont
  {Lewenstein}}, \bibinfo {author} {\bibfnamefont {M.~F.}\ \bibnamefont
  {Ciappina}}, \bibinfo {author} {\bibfnamefont {E.}~\bibnamefont {Pisanty}},
  \bibinfo {author} {\bibfnamefont {J.}~\bibnamefont {Rivera-Dean}}, \bibinfo
  {author} {\bibfnamefont {P.}~\bibnamefont {Stammer}}, \bibinfo {author}
  {\bibfnamefont {T.}~\bibnamefont {Lamprou}},\ and\ \bibinfo {author}
  {\bibfnamefont {P.}~\bibnamefont {Tzallas}},\ }\bibfield  {title} {\bibinfo
  {title} {Generation of optical {Schrödinger} cat states in intense
  laser–matter interactions},\ }\href
  {https://doi.org/10.1038/s41567-021-01317-w} {\bibfield  {journal} {\bibinfo
  {journal} {Nat. Phys.}\ }\textbf {\bibinfo {volume} {17}},\ \bibinfo {pages}
  {1104} (\bibinfo {year} {2021})}\BibitemShut {NoStop}%
\bibitem [{\citenamefont {Rivera-Dean}\ \emph {et~al.}(2022)\citenamefont
  {Rivera-Dean}, \citenamefont {Lamprou}, \citenamefont {Pisanty},
  \citenamefont {Stammer}, \citenamefont {Ordóñez}, \citenamefont {Maxwell},
  \citenamefont {Ciappina}, \citenamefont {Lewenstein},\ and\ \citenamefont
  {Tzallas}}]{rivera-dean_strong_2022}%
  \BibitemOpen
  \bibfield  {author} {\bibinfo {author} {\bibfnamefont {J.}~\bibnamefont
  {Rivera-Dean}}, \bibinfo {author} {\bibfnamefont {T.}~\bibnamefont
  {Lamprou}}, \bibinfo {author} {\bibfnamefont {E.}~\bibnamefont {Pisanty}},
  \bibinfo {author} {\bibfnamefont {P.}~\bibnamefont {Stammer}}, \bibinfo
  {author} {\bibfnamefont {A.}~\bibnamefont {Ordóñez}}, \bibinfo {author}
  {\bibfnamefont {A.~S.}\ \bibnamefont {Maxwell}}, \bibinfo {author}
  {\bibfnamefont {M.~F.}\ \bibnamefont {Ciappina}}, \bibinfo {author}
  {\bibfnamefont {M.}~\bibnamefont {Lewenstein}},\ and\ \bibinfo {author}
  {\bibfnamefont {P.}~\bibnamefont {Tzallas}},\ }\bibfield  {title} {\bibinfo
  {title} {Strong laser fields and their power to generate controllable
  high-photon-number coherent-state superpositions},\ }\href
  {https://doi.org/10.1103/PhysRevA.105.033714} {\bibfield  {journal} {\bibinfo
   {journal} {Phys. Rev. A}\ }\textbf {\bibinfo {volume} {105}},\ \bibinfo
  {pages} {033714} (\bibinfo {year} {2022})}\BibitemShut {NoStop}%
\bibitem [{\citenamefont {Stammer}(2022)}]{stammer_theory_2022}%
  \BibitemOpen
  \bibfield  {author} {\bibinfo {author} {\bibfnamefont {P.}~\bibnamefont
  {Stammer}},\ }\bibfield  {title} {\bibinfo {title} {Theory of entanglement
  and measurement in high-order harmonic generation},\ }\href
  {https://doi.org/10.1103/PhysRevA.106.L050402} {\bibfield  {journal}
  {\bibinfo  {journal} {Phys. Rev. A}\ }\textbf {\bibinfo {volume} {106}},\
  \bibinfo {pages} {L050402} (\bibinfo {year} {2022})}\BibitemShut {NoStop}%
\bibitem [{\citenamefont {Stammer}\ \emph {et~al.}(2022)\citenamefont
  {Stammer}, \citenamefont {Rivera-Dean}, \citenamefont {Lamprou},
  \citenamefont {Pisanty}, \citenamefont {Ciappina}, \citenamefont {Tzallas},\
  and\ \citenamefont {Lewenstein}}]{stammer_high_2022}%
  \BibitemOpen
  \bibfield  {author} {\bibinfo {author} {\bibfnamefont {P.}~\bibnamefont
  {Stammer}}, \bibinfo {author} {\bibfnamefont {J.}~\bibnamefont
  {Rivera-Dean}}, \bibinfo {author} {\bibfnamefont {T.}~\bibnamefont
  {Lamprou}}, \bibinfo {author} {\bibfnamefont {E.}~\bibnamefont {Pisanty}},
  \bibinfo {author} {\bibfnamefont {M.~F.}\ \bibnamefont {Ciappina}}, \bibinfo
  {author} {\bibfnamefont {P.}~\bibnamefont {Tzallas}},\ and\ \bibinfo {author}
  {\bibfnamefont {M.}~\bibnamefont {Lewenstein}},\ }\bibfield  {title}
  {\bibinfo {title} {High {Photon} {Number} {Entangled} {States} and {Coherent}
  {State} {Superposition} from the {Extreme} {Ultraviolet} to the {Far}
  {Infrared}},\ }\href {https://doi.org/10.1103/PhysRevLett.128.123603}
  {\bibfield  {journal} {\bibinfo  {journal} {Phys. Rev. Lett.}\ }\textbf
  {\bibinfo {volume} {128}},\ \bibinfo {pages} {123603} (\bibinfo {year}
  {2022})}\BibitemShut {NoStop}%
\bibitem [{\citenamefont {{\relax{J. Rivera-Dean, P. Stammer, A. S. Maxwell,
  Th. Lamprou, P. Tzallas, M. Lewenstein, and M. F.
  Ciappina}}}(2022)}]{Javier_PRA-ATI_2022}%
  \BibitemOpen
  \bibfield  {author} {\bibinfo {author} {\bibnamefont {{\relax{J. Rivera-Dean,
  P. Stammer, A. S. Maxwell, Th. Lamprou, P. Tzallas, M. Lewenstein, and M. F.
  Ciappina}}}},\ }\bibfield  {title} {\bibinfo {title} {\relax{Light-matter
  entanglement after above-threshold ionization processes in atoms}},\ }\href
  {https://doi.org/10.1103/PhysRevA.106.063705} {\bibfield  {journal} {\bibinfo
   {journal} {Phys. Rev. A}\ }\textbf {\bibinfo {volume} {106}},\ \bibinfo
  {pages} {063705} (\bibinfo {year} {2022})}\BibitemShut {NoStop}%
\bibitem [{\citenamefont {{\relax{J. Rivera-Dean, P. Stammer, A. S. Maxwell,
  Th. Lamprou, E. Pisanty, P. Tzallas, M. Lewenstein, and M. F.
  Ciappina}}}(2024)}]{Javier_PRA-H2_2024}%
  \BibitemOpen
  \bibfield  {author} {\bibinfo {author} {\bibnamefont {{\relax{J. Rivera-Dean,
  P. Stammer, A. S. Maxwell, Th. Lamprou, E. Pisanty, P. Tzallas, M.
  Lewenstein, and M. F. Ciappina}}}},\ }\bibfield  {title} {\bibinfo {title}
  {\relax{Quantum-optical analysis of high-order harmonic generation in
  H$^{+}_{2}$ molecules}},\ }\href
  {https://doi.org/10.1103/PhysRevA.109.033706} {\bibfield  {journal} {\bibinfo
   {journal} {Phys. Rev. A}\ }\textbf {\bibinfo {volume} {109}},\ \bibinfo
  {pages} {033706} (\bibinfo {year} {2024})}\BibitemShut {NoStop}%
\bibitem [{\citenamefont {\relax{Á. Gombkötő, S. Varró, B. G. Pusztai, I.
  Magashegyi, A. Czirják, S. Hack, and P. Földi}}(2024)}]{Foldi_PRA_2024}%
  \BibitemOpen
  \bibfield  {author} {\bibinfo {author} {\bibnamefont {\relax{Á. Gombkötő,
  S. Varró, B. G. Pusztai, I. Magashegyi, A. Czirják, S. Hack, and P.
  Földi}}},\ }\bibfield  {title} {\bibinfo {title} {Parametric model for
  high-order harmonic generation with quantized fields},\ }\href
  {https://doi.org/doi.org/10.1103/PhysRevA.109.053717} {\bibfield  {journal}
  {\bibinfo  {journal} {Phys. Rev. A}\ }\textbf {\bibinfo {volume} {109}},\
  \bibinfo {pages} {053717} (\bibinfo {year} {2024})}\BibitemShut {NoStop}%
\bibitem [{\citenamefont {Rivera-Dean}\ \emph
  {et~al.}(2024{\natexlab{a}})\citenamefont {Rivera-Dean}, \citenamefont
  {Stammer}, \citenamefont {Maxwell}, \citenamefont {Lamprou}, \citenamefont
  {Ordóñez}, \citenamefont {Pisanty}, \citenamefont {Tzallas}, \citenamefont
  {Lewenstein},\ and\ \citenamefont {Ciappina}}]{Javier_PRB_2024}%
  \BibitemOpen
  \bibfield  {author} {\bibinfo {author} {\bibfnamefont {J.}~\bibnamefont
  {Rivera-Dean}}, \bibinfo {author} {\bibfnamefont {P.}~\bibnamefont
  {Stammer}}, \bibinfo {author} {\bibfnamefont {A.~S.}\ \bibnamefont
  {Maxwell}}, \bibinfo {author} {\bibfnamefont {T.}~\bibnamefont {Lamprou}},
  \bibinfo {author} {\bibfnamefont {A.~F.}\ \bibnamefont {Ordóñez}}, \bibinfo
  {author} {\bibfnamefont {E.}~\bibnamefont {Pisanty}}, \bibinfo {author}
  {\bibfnamefont {P.}~\bibnamefont {Tzallas}}, \bibinfo {author} {\bibfnamefont
  {M.}~\bibnamefont {Lewenstein}},\ and\ \bibinfo {author} {\bibfnamefont
  {M.~F.}\ \bibnamefont {Ciappina}},\ }\bibfield  {title} {\bibinfo {title}
  {Nonclassical states of light after high-harmonic generation in
  semiconductors: A bloch-based perspective},\ }\href@noop {} {\bibfield
  {journal} {\bibinfo  {journal} {Phys. Rev. B}\ }\textbf {\bibinfo {volume}
  {109}},\ \bibinfo {pages} {035203} (\bibinfo {year}
  {2024}{\natexlab{a}})}\BibitemShut {NoStop}%
\bibitem [{\citenamefont {Gonoskov}\ \emph {et~al.}(2024)\citenamefont
  {Gonoskov}, \citenamefont {Sondenheimer}, \citenamefont {Hünecke},
  \citenamefont {Kartashov}, \citenamefont {Peschel},\ and\ \citenamefont
  {Gräfe}}]{Gonoskov_PRB_2024}%
  \BibitemOpen
  \bibfield  {author} {\bibinfo {author} {\bibfnamefont {I.}~\bibnamefont
  {Gonoskov}}, \bibinfo {author} {\bibfnamefont {R.}~\bibnamefont
  {Sondenheimer}}, \bibinfo {author} {\bibfnamefont {C.}~\bibnamefont
  {Hünecke}}, \bibinfo {author} {\bibfnamefont {D.}~\bibnamefont {Kartashov}},
  \bibinfo {author} {\bibfnamefont {U.}~\bibnamefont {Peschel}},\ and\ \bibinfo
  {author} {\bibfnamefont {S.}~\bibnamefont {Gräfe}},\ }\bibfield  {title}
  {\bibinfo {title} {Nonclassical light generation and control from
  laser-driven semiconductor intraband excitations},\ }\href@noop {} {\bibfield
   {journal} {\bibinfo  {journal} {Phys. Rev. B}\ }\textbf {\bibinfo {volume}
  {109}},\ \bibinfo {pages} {125110} (\bibinfo {year} {2024})}\BibitemShut
  {NoStop}%
\bibitem [{\citenamefont {{\relax{Th. Lamprou, J. Rivera-Dean, P. Stammer, M.
  Lewenstein, and P. Tzallas}}}(2025)}]{Lamprou_PRL-Nonlinear_2025}%
  \BibitemOpen
  \bibfield  {author} {\bibinfo {author} {\bibnamefont {{\relax{Th. Lamprou, J.
  Rivera-Dean, P. Stammer, M. Lewenstein, and P. Tzallas}}}},\ }\bibfield
  {title} {\bibinfo {title} {\relax{Nonlinear optics using intense optical
  coherent state superpositions}},\ }\href
  {https://doi.org/10.1103/PhysRevLett.134.013601} {\bibfield  {journal}
  {\bibinfo  {journal} {Phys. Rev. Lett.}\ }\textbf {\bibinfo {volume} {134}},\
  \bibinfo {pages} {013601} (\bibinfo {year} {2025})}\BibitemShut {NoStop}%
\bibitem [{\citenamefont {\relax{C. S. Lange, T. Hansen, and L. B.
  Madsen}}(2025)}]{Madsen_PRL_2025}%
  \BibitemOpen
  \bibfield  {author} {\bibinfo {author} {\bibnamefont {\relax{C. S. Lange, T.
  Hansen, and L. B. Madsen}}},\ }\bibfield  {title} {\bibinfo {title}
  {Excitonic enhancement of squeezed light in quantum-optical high-harmonic
  generation from a mott insulator},\ }\href
  {https://doi.org/10.1103/wyk5-k8tk} {\bibfield  {journal} {\bibinfo
  {journal} {Phys. Rev. Lett.}\ }\textbf {\bibinfo {volume} {135}},\ \bibinfo
  {pages} {043603} (\bibinfo {year} {2025})}\BibitemShut {NoStop}%
\bibitem [{\citenamefont {{\relax{S. Yi, N. D. Klimkin, G. G. Brown, O.
  Smirnova, S. Patchkovskii, I. Babushkin, and M.
  Ivanov}}}(2025)}]{Yi_PRX_2025}%
  \BibitemOpen
  \bibfield  {author} {\bibinfo {author} {\bibnamefont {{\relax{S. Yi, N. D.
  Klimkin, G. G. Brown, O. Smirnova, S. Patchkovskii, I. Babushkin, and M.
  Ivanov}}}},\ }\bibfield  {title} {\bibinfo {title} {\relax{Generation of
  massively entangled bright states of light during harmonic generation in
  resonant media}},\ }\href {https://doi.org/10.1103/PhysRevX.15.011023}
  {\bibfield  {journal} {\bibinfo  {journal} {Phys. Rev. X}\ }\textbf {\bibinfo
  {volume} {15}},\ \bibinfo {pages} {011023} (\bibinfo {year}
  {2025})}\BibitemShut {NoStop}%
\bibitem [{\citenamefont {{\relax{A. Pizzi, A. Gorlach, N. Rivera, A.
  Nunnenkamp, and I. Kaminer}}}(2023)}]{Pizzi_NatPhys-CorrAtoms_2023}%
  \BibitemOpen
  \bibfield  {author} {\bibinfo {author} {\bibnamefont {{\relax{A. Pizzi, A.
  Gorlach, N. Rivera, A. Nunnenkamp, and I. Kaminer}}}},\ }\bibfield  {title}
  {\bibinfo {title} {\relax{Light emission from strongly driven manybody
  systems}},\ }\href {https://doi.org/10.1038/s41567-022-01910-7} {\bibfield
  {journal} {\bibinfo  {journal} {Nat. Phys.}\ }\textbf {\bibinfo {volume}
  {19}},\ \bibinfo {pages} {551–561} (\bibinfo {year} {2023})}\BibitemShut
  {NoStop}%
\bibitem [{\citenamefont {{\relax{E. S. Andrianov and O. I.
  Tolstikhin}}}(2024)}]{Andrianov_PRA-DenseMedia_2024}%
  \BibitemOpen
  \bibfield  {author} {\bibinfo {author} {\bibnamefont {{\relax{E. S. Andrianov
  and O. I. Tolstikhin}}}},\ }\bibfield  {title} {\bibinfo {title}
  {\relax{Formation of nonclassical and non-gaussian states of a strong
  electromagnetic field due to its interaction with free electrons produced by
  ionization of a target gas}},\ }\href
  {https://doi.org/10.1103/PhysRevA.110.023115} {\bibfield  {journal} {\bibinfo
   {journal} {Phys. Rev. A}\ }\textbf {\bibinfo {volume} {110}},\ \bibinfo
  {pages} {023115} (\bibinfo {year} {2024})}\BibitemShut {NoStop}%
\bibitem [{\citenamefont {{\relax{C. S. Lange, T. Hansen, and L. B.
  Madsen}}}(2024)}]{Lange_PRA-ElectCorr_2024}%
  \BibitemOpen
  \bibfield  {author} {\bibinfo {author} {\bibnamefont {{\relax{C. S. Lange, T.
  Hansen, and L. B. Madsen}}}},\ }\bibfield  {title} {\bibinfo {title}
  {\relax{Electron correlation-induced nonclassicality of light from high order
  harmonic generation}},\ }\href {https://doi.org/10.1103/PhysRevA.109.033110}
  {\bibfield  {journal} {\bibinfo  {journal} {Phys. Rev. A}\ }\textbf {\bibinfo
  {volume} {109}},\ \bibinfo {pages} {033110} (\bibinfo {year}
  {2024})}\BibitemShut {NoStop}%
\bibitem [{\citenamefont {{\relax{D. Theidel, V. Cotte, R. Sondenheimer, V.
  Shiriaeva, M. Froidevaux, V. Severin, P. Mosel, A. MerdjiLarue, S. Fröhlich,
  K.-A. Weber, U. Morgner, M. Kovacev, J. Biegert, and H.
  Merdji}}}(2024)}]{Theidel_PRXQuantum_2024}%
  \BibitemOpen
  \bibfield  {author} {\bibinfo {author} {\bibnamefont {{\relax{D. Theidel, V.
  Cotte, R. Sondenheimer, V. Shiriaeva, M. Froidevaux, V. Severin, P. Mosel, A.
  MerdjiLarue, S. Fröhlich, K.-A. Weber, U. Morgner, M. Kovacev, J. Biegert,
  and H. Merdji}}}},\ }\bibfield  {title} {\bibinfo {title} {\relax{Evidence of
  the quantum optical nature of high-harmonic generation}},\ }\href
  {https://doi.org/10.1103/PRXQuantum.5.040319} {\bibfield  {journal} {\bibinfo
   {journal} {PRX Quantum}\ }\textbf {\bibinfo {volume} {5}},\ \bibinfo {pages}
  {040319} (\bibinfo {year} {2024})}\BibitemShut {NoStop}%
\bibitem [{\citenamefont {\relax{E. S. Andrianov and O. I.
  Tolstikhin}}(2025)}]{Adrianov_PRA_2025}%
  \BibitemOpen
  \bibfield  {author} {\bibinfo {author} {\bibnamefont {\relax{E. S. Andrianov
  and O. I. Tolstikhin}}},\ }\bibfield  {title} {\bibinfo {title} {Formation of
  a schrödinger cat state of a strong circularly polarized laser field due to
  thomson scattering by free electrons},\ }\href
  {https://doi.org/doi.org/10.1103/3gjp-f7br} {\bibfield  {journal} {\bibinfo
  {journal} {Phys. Rev. A}\ }\textbf {\bibinfo {volume} {112}},\ \bibinfo
  {pages} {013117} (\bibinfo {year} {2025})}\BibitemShut {NoStop}%
\bibitem [{\citenamefont {\relax{S. Imai, A. Ono, N.
  Tsuji}}(2025)}]{Imai_arXiv_2025}%
  \BibitemOpen
  \bibfield  {author} {\bibinfo {author} {\bibnamefont {\relax{S. Imai, A. Ono,
  N. Tsuji}}},\ }\bibfield  {title} {\bibinfo {title} {Electron dynamics
  induced by quantum cat-state light},\ }\href
  {https://doi.org/doi.org/10.48550/arXiv.2501.16801} {\bibfield  {journal}
  {\bibinfo  {journal} {arXiv:2501.16801}\ } (\bibinfo {year}
  {2025})}\BibitemShut {NoStop}%
\bibitem [{\citenamefont {Stammer}\ \emph {et~al.}(2023)\citenamefont
  {Stammer}, \citenamefont {Rivera-Dean}, \citenamefont {Maxwell},
  \citenamefont {Lamprou}, \citenamefont {Ordóñez}, \citenamefont {Ciappina},
  \citenamefont {Tzallas},\ and\ \citenamefont
  {Lewenstein}}]{stammer_quantum_2023}%
  \BibitemOpen
  \bibfield  {author} {\bibinfo {author} {\bibfnamefont {P.}~\bibnamefont
  {Stammer}}, \bibinfo {author} {\bibfnamefont {J.}~\bibnamefont
  {Rivera-Dean}}, \bibinfo {author} {\bibfnamefont {A.~S.}\ \bibnamefont
  {Maxwell}}, \bibinfo {author} {\bibfnamefont {T.}~\bibnamefont {Lamprou}},
  \bibinfo {author} {\bibfnamefont {A.}~\bibnamefont {Ordóñez}}, \bibinfo
  {author} {\bibfnamefont {M.~F.}\ \bibnamefont {Ciappina}}, \bibinfo {author}
  {\bibfnamefont {P.}~\bibnamefont {Tzallas}},\ and\ \bibinfo {author}
  {\bibfnamefont {M.}~\bibnamefont {Lewenstein}},\ }\bibfield  {title}
  {\bibinfo {title} {Quantum {Electrodynamics} of {Intense} {Laser}-{Matter}
  {Interactions}: {A} {Tool} for {Quantum} {State} {Engineering}},\ }\href
  {https://doi.org/10.1103/PRXQuantum.4.010201} {\bibfield  {journal} {\bibinfo
   {journal} {PRX Quantum}\ }\textbf {\bibinfo {volume} {4}},\ \bibinfo {pages}
  {010201} (\bibinfo {year} {2023})}\BibitemShut {NoStop}%
\bibitem [{\citenamefont {Bhattacharya}\ \emph {et~al.}(2023)\citenamefont
  {Bhattacharya}, \citenamefont {Lamprou}, \citenamefont {Maxwell},
  \citenamefont {Ordonez}, \citenamefont {Pisanty}, \citenamefont
  {Rivera-Dean}, \citenamefont {Stammer}, \citenamefont {Ciappina},
  \citenamefont {Lewenstein},\ and\ \citenamefont
  {Tzallas}}]{Bhattacharya_2023}%
  \BibitemOpen
  \bibfield  {author} {\bibinfo {author} {\bibfnamefont {U.}~\bibnamefont
  {Bhattacharya}}, \bibinfo {author} {\bibfnamefont {T.}~\bibnamefont
  {Lamprou}}, \bibinfo {author} {\bibfnamefont {A.~S.}\ \bibnamefont
  {Maxwell}}, \bibinfo {author} {\bibfnamefont {A.}~\bibnamefont {Ordonez}},
  \bibinfo {author} {\bibfnamefont {E.}~\bibnamefont {Pisanty}}, \bibinfo
  {author} {\bibfnamefont {J.}~\bibnamefont {Rivera-Dean}}, \bibinfo {author}
  {\bibfnamefont {P.}~\bibnamefont {Stammer}}, \bibinfo {author} {\bibfnamefont
  {M.~F.}\ \bibnamefont {Ciappina}}, \bibinfo {author} {\bibfnamefont
  {M.}~\bibnamefont {Lewenstein}},\ and\ \bibinfo {author} {\bibfnamefont
  {P.}~\bibnamefont {Tzallas}},\ }\bibfield  {title} {\bibinfo {title}
  {Strong–laser–field physics, non–classical light states and quantum
  information science.},\ }\href {https://doi.org/10.1088/1361-6633/acea31}
  {\bibfield  {journal} {\bibinfo  {journal} {Rep. Prog. Phys.}\ }\textbf
  {\bibinfo {volume} {86}},\ \bibinfo {pages} {094401} (\bibinfo {year}
  {2023})}\BibitemShut {NoStop}%
\bibitem [{\citenamefont {Cruz-Rodriguez}\ \emph {et~al.}(2024)\citenamefont
  {Cruz-Rodriguez}, \citenamefont {Dey}, \citenamefont {Freibert},\ and\
  \citenamefont {Stammer}}]{cruz-rodriguez_quantum_2024}%
  \BibitemOpen
  \bibfield  {author} {\bibinfo {author} {\bibfnamefont {L.}~\bibnamefont
  {Cruz-Rodriguez}}, \bibinfo {author} {\bibfnamefont {D.}~\bibnamefont {Dey}},
  \bibinfo {author} {\bibfnamefont {A.}~\bibnamefont {Freibert}},\ and\
  \bibinfo {author} {\bibfnamefont {P.}~\bibnamefont {Stammer}},\ }\bibfield
  {title} {\bibinfo {title} {Quantum phenomena in attosecond science},\ }\href
  {https://doi.org/10.1038/s42254-024-00769-2} {\bibfield  {journal} {\bibinfo
  {journal} {Nature Reviews Physics}\ }\textbf {\bibinfo {volume} {6}},\
  \bibinfo {pages} {691} (\bibinfo {year} {2024})}\BibitemShut {NoStop}%
\bibitem [{\citenamefont {{\relax{Th. Lamprou, P. Stammer, J. Rivera-Dean, N.
  Tsatrafyllis, M. F. Ciappina, M. Lewenstein, and P.
  Tzallas}}}(2025)}]{Lamprou_JPB-Review_2025}%
  \BibitemOpen
  \bibfield  {author} {\bibinfo {author} {\bibnamefont {{\relax{Th. Lamprou, P.
  Stammer, J. Rivera-Dean, N. Tsatrafyllis, M. F. Ciappina, M. Lewenstein, and
  P. Tzallas}}}},\ }\bibfield  {title} {\bibinfo {title} {\relax{Recent
  developments in the generation of non–classical and entangled light states
  using intense laser–matter interactions}},\ }\href
  {https://doi.org/10.1088/1361-6455/add9fe} {\bibfield  {journal} {\bibinfo
  {journal} {J. Phys. B: At. Mol. Opt. Phys.}\ }\textbf {\bibinfo {volume}
  {58}},\ \bibinfo {pages} {132001} (\bibinfo {year} {2025})}\BibitemShut
  {NoStop}%
\bibitem [{\citenamefont {{\relax{J. Heimerl, A. Mikhaylov, S. Meier, H.
  Höllerer, I. Kaminer, M. Chekhova, and P.
  Hommelhoff}}}(2024)}]{Heimerl_NatPhys-MultiElectrn_2024}%
  \BibitemOpen
  \bibfield  {author} {\bibinfo {author} {\bibnamefont {{\relax{J. Heimerl, A.
  Mikhaylov, S. Meier, H. Höllerer, I. Kaminer, M. Chekhova, and P.
  Hommelhoff}}}},\ }\bibfield  {title} {\bibinfo {title} {\relax{Multiphoton
  electron emission with non-classical light}},\ }\href {https://doi.org/...}
  {\bibfield  {journal} {\bibinfo  {journal} {Nat. Phys.}\ }\textbf {\bibinfo
  {volume} {20}},\ \bibinfo {pages} {945} (\bibinfo {year} {2024})}\BibitemShut
  {NoStop}%
\bibitem [{\citenamefont {Heimerl}\ \emph {et~al.}(2025)\citenamefont
  {Heimerl}, \citenamefont {Rasputnyi}, \citenamefont {Pölloth}, \citenamefont
  {Meier}, \citenamefont {Chekhova},\ and\ \citenamefont
  {Hommelhoff}}]{heimerl_driving_2025}%
  \BibitemOpen
  \bibfield  {author} {\bibinfo {author} {\bibfnamefont {J.}~\bibnamefont
  {Heimerl}}, \bibinfo {author} {\bibfnamefont {A.}~\bibnamefont {Rasputnyi}},
  \bibinfo {author} {\bibfnamefont {J.}~\bibnamefont {Pölloth}}, \bibinfo
  {author} {\bibfnamefont {S.}~\bibnamefont {Meier}}, \bibinfo {author}
  {\bibfnamefont {M.}~\bibnamefont {Chekhova}},\ and\ \bibinfo {author}
  {\bibfnamefont {P.}~\bibnamefont {Hommelhoff}},\ }\href
  {https://doi.org/10.48550/arXiv.2503.22464} {\bibinfo {title} {Driving
  electrons at needle tips strongly with quantum light}} (\bibinfo {year}
  {2025}),\ \bibinfo {note} {arXiv:2503.22464 [quant-ph]}\BibitemShut {NoStop}%
\bibitem [{\citenamefont {{\relax{H. Liu, H. Zhang, X. Wang, and J.
  Yuan}}}(2025)}]{Liu_PRL-AtomIon_2025}%
  \BibitemOpen
  \bibfield  {author} {\bibinfo {author} {\bibnamefont {{\relax{H. Liu, H.
  Zhang, X. Wang, and J. Yuan}}}},\ }\bibfield  {title} {\bibinfo {title}
  {\relax{Atomic double ionization with quantum light}},\ }\href
  {https://doi.org/10.1103/PhysRevLett.134.123202} {\bibfield  {journal}
  {\bibinfo  {journal} {Phys. Rev. Lett.}\ }\textbf {\bibinfo {volume} {134}},\
  \bibinfo {pages} {123202} (\bibinfo {year} {2025})}\BibitemShut {NoStop}%
\bibitem [{\citenamefont {\relax{Z. Lyu, F. Sun, Y. Fang, Q. He and Y.
  Liu}}(2025)}]{Lyu_PRR_2025}%
  \BibitemOpen
  \bibfield  {author} {\bibinfo {author} {\bibnamefont {\relax{Z. Lyu, F. Sun,
  Y. Fang, Q. He and Y. Liu}}},\ }\bibfield  {title} {\bibinfo {title} {Effect
  of photon quantum statistics on electrons in above-threshold ionization},\
  }\href {https://doi.org/10.1103/PhysRevResearch.7.L012072} {\bibfield
  {journal} {\bibinfo  {journal} {Phys. Rev. Res.}\ }\textbf {\bibinfo {volume}
  {7}},\ \bibinfo {pages} {L012072} (\bibinfo {year} {2025})}\BibitemShut
  {NoStop}%
\bibitem [{\citenamefont {{\relax{G. Mouloudakis and P.
  Lambropoulos}}}(2020)}]{Mouloudakis_PRA-Spectr_2019}%
  \BibitemOpen
  \bibfield  {author} {\bibinfo {author} {\bibnamefont {{\relax{G. Mouloudakis
  and P. Lambropoulos}}}},\ }\bibfield  {title} {\bibinfo {title}
  {\relax{Pairing superbunching with compounded nonlinearity in a resonant
  transition}},\ }\href {https://doi.org/...} {\bibfield  {journal} {\bibinfo
  {journal} {Phys. Rev. A}\ }\textbf {\bibinfo {volume} {102}},\ \bibinfo
  {pages} {023713} (\bibinfo {year} {2020})}\BibitemShut {NoStop}%
\bibitem [{\citenamefont {{\relax{A. Gorlach, M. E. Tzur, M. Birk, et
  al.,}}}(2023)}]{Gorlach_NatPhys-HHG-BSV_2023}%
  \BibitemOpen
  \bibfield  {author} {\bibinfo {author} {\bibnamefont {{\relax{A. Gorlach, M.
  E. Tzur, M. Birk, et al.,}}}},\ }\bibfield  {title} {\bibinfo {title}
  {\relax{High-harmonic generation driven by quantum light}},\ }\href
  {https://doi.org/10.1038/s41567-023-02127-y} {\bibfield  {journal} {\bibinfo
  {journal} {Nat. Phys.}\ }\textbf {\bibinfo {volume} {19}},\ \bibinfo {pages}
  {1689} (\bibinfo {year} {2023})}\BibitemShut {NoStop}%
\bibitem [{\citenamefont {{\relax{S. Wang, S. Yu, X. Lai, and X.
  Liu}}}(2024)}]{Wang_PRR-HHG-BSV_2024}%
  \BibitemOpen
  \bibfield  {author} {\bibinfo {author} {\bibnamefont {{\relax{S. Wang, S. Yu,
  X. Lai, and X. Liu}}}},\ }\bibfield  {title} {\bibinfo {title} {\relax{High
  harmonic generation from an atom in a squeezed-vacuum environment}},\ }\href
  {https://doi.org/10.1103/PhysRevResearch.6.033010} {\bibfield  {journal}
  {\bibinfo  {journal} {Phys. Rev. Res.}\ }\textbf {\bibinfo {volume} {6}},\
  \bibinfo {pages} {033010} (\bibinfo {year} {2024})}\BibitemShut {NoStop}%
\bibitem [{\citenamefont {{\relax{P.
  Stammer}}}(2024)}]{Stammer_NatPhys-HHG-BSV_2024}%
  \BibitemOpen
  \bibfield  {author} {\bibinfo {author} {\bibnamefont {{\relax{P.
  Stammer}}}},\ }\bibfield  {title} {\bibinfo {title} {\relax{On the
  limitations of the semi-classical picture in high harmonic generation}},\
  }\href {https://doi.org/10.1038/s41567-024-02579-w} {\bibfield  {journal}
  {\bibinfo  {journal} {Nat. Phys.}\ }\textbf {\bibinfo {volume} {20}},\
  \bibinfo {pages} {1040} (\bibinfo {year} {2024})}\BibitemShut {NoStop}%
\bibitem [{\citenamefont {{\relax{M. E. Tzur, M. Birk, A. Gorlach, I. Kaminer,
  M. Krüger, and O. Cohen}}}(2024)}]{Tzur_PRR-SqeezedHHG_2024}%
  \BibitemOpen
  \bibfield  {author} {\bibinfo {author} {\bibnamefont {{\relax{M. E. Tzur, M.
  Birk, A. Gorlach, I. Kaminer, M. Krüger, and O. Cohen}}}},\ }\bibfield
  {title} {\bibinfo {title} {\relax{Generation of squeezed high-order
  harmonics}},\ }\href {https://doi.org/10.1103/PhysRevResearch.6.033079}
  {\bibfield  {journal} {\bibinfo  {journal} {Phys. Rev. Res.}\ }\textbf
  {\bibinfo {volume} {6}},\ \bibinfo {pages} {033079} (\bibinfo {year}
  {2024})}\BibitemShut {NoStop}%
\bibitem [{\citenamefont {\relax{J. Rivera-Dean, P. Stammer, M. F. Ciappina,
  and M. Lewenstein}}(2025)}]{Javier_PRL-BSV_2025}%
  \BibitemOpen
  \bibfield  {author} {\bibinfo {author} {\bibnamefont {\relax{J. Rivera-Dean,
  P. Stammer, M. F. Ciappina, and M. Lewenstein}}},\ }\bibfield  {title}
  {\bibinfo {title} {Structured squeezed light allows for high-harmonic
  generation in classical forbidden geometries},\ }\href
  {https://doi.org/10.1103/4hdl-bdwj} {\bibfield  {journal} {\bibinfo
  {journal} {Phys. Rev. Lett.}\ }\textbf {\bibinfo {volume} {135}},\ \bibinfo
  {pages} {013801} (\bibinfo {year} {2025})}\BibitemShut {NoStop}%
\bibitem [{\citenamefont {\relax{S. Lemieux, S. A. Jalil, D. Purschke, N.
  Boroumand, T. J. Hammond, D. Villeneuve, A. Naumov, T. Brabec, and G.
  Vampa}}(2025)}]{Lemieux_NatPhoton-BSV_2025}%
  \BibitemOpen
  \bibfield  {author} {\bibinfo {author} {\bibnamefont {\relax{S. Lemieux, S.
  A. Jalil, D. Purschke, N. Boroumand, T. J. Hammond, D. Villeneuve, A. Naumov,
  T. Brabec, and G. Vampa}}},\ }\bibfield  {title} {\bibinfo {title} {Photon
  bunching in high-harmonic emission controlled by quantum light},\ }\href
  {https://doi.org/10.1038/s41566-025-01673-6} {\bibfield  {journal} {\bibinfo
  {journal} {Nat. Photon.}\ }\textbf {\bibinfo {volume} {19}},\ \bibinfo
  {pages} {767–771} (\bibinfo {year} {2025})}\BibitemShut {NoStop}%
\bibitem [{\citenamefont {\relax{A. Rasputnyi, Z. Chen, M. Birk, O. Cohen, I.
  Kaminer, M. Krüger, D. Seletskiy, M. Chekhova, and F.
  Tani}}(2024)}]{Rasputnyi_NatPhys-BSV_2024}%
  \BibitemOpen
  \bibfield  {author} {\bibinfo {author} {\bibnamefont {\relax{A. Rasputnyi, Z.
  Chen, M. Birk, O. Cohen, I. Kaminer, M. Krüger, D. Seletskiy, M. Chekhova,
  and F. Tani}}},\ }\bibfield  {title} {\bibinfo {title} {High-harmonic
  generation by a bright squeezed vacuum},\ }\href
  {https://doi.org/10.1038/s41567-024-02659-x} {\bibfield  {journal} {\bibinfo
  {journal} {Nat. Phys.}\ }\textbf {\bibinfo {volume} {20}},\ \bibinfo {pages}
  {1960} (\bibinfo {year} {2024})}\BibitemShut {NoStop}%
\bibitem [{\citenamefont {Gothelf}\ \emph {et~al.}(2025)\citenamefont
  {Gothelf}, \citenamefont {Lange},\ and\ \citenamefont
  {Madsen}}]{gothelf2025high}%
  \BibitemOpen
  \bibfield  {author} {\bibinfo {author} {\bibfnamefont {R.~V.}\ \bibnamefont
  {Gothelf}}, \bibinfo {author} {\bibfnamefont {C.~S.}\ \bibnamefont {Lange}},\
  and\ \bibinfo {author} {\bibfnamefont {L.~B.}\ \bibnamefont {Madsen}},\
  }\bibfield  {title} {\bibinfo {title} {High-order harmonic generation in a
  crystal driven by quantum light},\ }\href
  {https://doi.org/10.1103/PhysRevA.111.063105} {\bibfield  {journal} {\bibinfo
   {journal} {Phys. Rev. A}\ }\textbf {\bibinfo {volume} {111}},\ \bibinfo
  {pages} {063105} (\bibinfo {year} {2025})}\BibitemShut {NoStop}%
\bibitem [{\citenamefont {Tzur}\ \emph {et~al.}(2025)\citenamefont {Tzur},
  \citenamefont {Mor}, \citenamefont {Yaffe}, \citenamefont {Birk},
  \citenamefont {Rasputnyi}, \citenamefont {Kneller}, \citenamefont {Nisim},
  \citenamefont {Kaminer}, \citenamefont {Krüger}, \citenamefont {Dudovich},\
  and\ \citenamefont {Cohen}}]{tzur_measuring_2025}%
  \BibitemOpen
  \bibfield  {author} {\bibinfo {author} {\bibfnamefont {M.~E.}\ \bibnamefont
  {Tzur}}, \bibinfo {author} {\bibfnamefont {C.}~\bibnamefont {Mor}}, \bibinfo
  {author} {\bibfnamefont {N.}~\bibnamefont {Yaffe}}, \bibinfo {author}
  {\bibfnamefont {M.}~\bibnamefont {Birk}}, \bibinfo {author} {\bibfnamefont
  {A.}~\bibnamefont {Rasputnyi}}, \bibinfo {author} {\bibfnamefont
  {O.}~\bibnamefont {Kneller}}, \bibinfo {author} {\bibfnamefont
  {I.}~\bibnamefont {Nisim}}, \bibinfo {author} {\bibfnamefont
  {I.}~\bibnamefont {Kaminer}}, \bibinfo {author} {\bibfnamefont
  {M.}~\bibnamefont {Krüger}}, \bibinfo {author} {\bibfnamefont
  {N.}~\bibnamefont {Dudovich}},\ and\ \bibinfo {author} {\bibfnamefont
  {O.}~\bibnamefont {Cohen}},\ }\href
  {https://doi.org/10.48550/arXiv.2502.09427} {\bibinfo {title} {Measuring and
  controlling the birth of quantum attosecond pulses}} (\bibinfo {year}
  {2025}),\ \bibinfo {note} {arXiv:2502.09427 [physics]}\BibitemShut {NoStop}%
\bibitem [{\citenamefont {Stammer}\ \emph {et~al.}(2025)\citenamefont
  {Stammer}, \citenamefont {Rivera-Dean}, \citenamefont {Ciappina},\ and\
  \citenamefont {Lewenstein}}]{stammer_weak_2025}%
  \BibitemOpen
  \bibfield  {author} {\bibinfo {author} {\bibfnamefont {P.}~\bibnamefont
  {Stammer}}, \bibinfo {author} {\bibfnamefont {J.}~\bibnamefont
  {Rivera-Dean}}, \bibinfo {author} {\bibfnamefont {M.~F.}\ \bibnamefont
  {Ciappina}},\ and\ \bibinfo {author} {\bibfnamefont {M.}~\bibnamefont
  {Lewenstein}},\ }\href {https://doi.org/10.48550/arXiv.2508.09048} {\bibinfo
  {title} {Weak measurement in strong laser field physics}} (\bibinfo {year}
  {2025}),\ \bibinfo {note} {arXiv:2508.09048 [quant-ph]}\BibitemShut {NoStop}%
\bibitem [{\citenamefont {\relax{P. Stammer, J. Rivera-Dean, A. S. Maxwell, Th.
  Lamprou, J. Argüello-Luengo, P. Tzallas, M. F. Ciappina, and M.
  Lewenstein}}(2024)}]{Stammer_PRL-Squeezing_2024}%
  \BibitemOpen
  \bibfield  {author} {\bibinfo {author} {\bibnamefont {\relax{P. Stammer, J.
  Rivera-Dean, A. S. Maxwell, Th. Lamprou, J. Argüello-Luengo, P. Tzallas, M.
  F. Ciappina, and M. Lewenstein}}},\ }\bibfield  {title} {\bibinfo {title}
  {Entanglement and squeezing of the optical field modes in high harmonic
  generation},\ }\href {https://doi.org/10.1103/PhysRevLett.132.143603}
  {\bibfield  {journal} {\bibinfo  {journal} {Phys. Rev. Lett.}\ }\textbf
  {\bibinfo {volume} {132}},\ \bibinfo {pages} {143603} (\bibinfo {year}
  {2024})}\BibitemShut {NoStop}%
\bibitem [{\citenamefont {Rivera-Dean}\ \emph
  {et~al.}(2024{\natexlab{b}})\citenamefont {Rivera-Dean}, \citenamefont
  {Crispin}, \citenamefont {Stammer}, \citenamefont {Lamprou}, \citenamefont
  {Pisanty}, \citenamefont {Krüger}, \citenamefont {Tzallas}, \citenamefont
  {Lewenstein},\ and\ \citenamefont {Ciappina}}]{rivera-dean_squeezed_2024}%
  \BibitemOpen
  \bibfield  {author} {\bibinfo {author} {\bibfnamefont {J.}~\bibnamefont
  {Rivera-Dean}}, \bibinfo {author} {\bibfnamefont {H.~B.}\ \bibnamefont
  {Crispin}}, \bibinfo {author} {\bibfnamefont {P.}~\bibnamefont {Stammer}},
  \bibinfo {author} {\bibfnamefont {T.}~\bibnamefont {Lamprou}}, \bibinfo
  {author} {\bibfnamefont {E.}~\bibnamefont {Pisanty}}, \bibinfo {author}
  {\bibfnamefont {M.}~\bibnamefont {Krüger}}, \bibinfo {author} {\bibfnamefont
  {P.}~\bibnamefont {Tzallas}}, \bibinfo {author} {\bibfnamefont
  {M.}~\bibnamefont {Lewenstein}},\ and\ \bibinfo {author} {\bibfnamefont
  {M.~F.}\ \bibnamefont {Ciappina}},\ }\bibfield  {title} {\bibinfo {title}
  {Squeezed states of light after high-order harmonic generation in excited
  atomic systems},\ }\href {https://doi.org/10.1103/PhysRevA.110.063118}
  {\bibfield  {journal} {\bibinfo  {journal} {Physical Review A}\ }\textbf
  {\bibinfo {volume} {110}},\ \bibinfo {pages} {063118} (\bibinfo {year}
  {2024}{\natexlab{b}})}\BibitemShut {NoStop}%
\bibitem [{\citenamefont {Lange}\ and\ \citenamefont
  {Madsen}(2025)}]{lange_hierarchy_2025}%
  \BibitemOpen
  \bibfield  {author} {\bibinfo {author} {\bibfnamefont {C.~S.}\ \bibnamefont
  {Lange}}\ and\ \bibinfo {author} {\bibfnamefont {L.~B.}\ \bibnamefont
  {Madsen}},\ }\bibfield  {title} {\bibinfo {title} {Hierarchy of
  approximations for describing quantum light from high-harmonic generation:
  {A} {Fermi}-{Hubbard}-model study},\ }\href
  {https://doi.org/10.1103/PhysRevA.111.013113} {\bibfield  {journal} {\bibinfo
   {journal} {Physical Review A}\ }\textbf {\bibinfo {volume} {111}},\ \bibinfo
  {pages} {013113} (\bibinfo {year} {2025})}\BibitemShut {NoStop}%
\bibitem [{\citenamefont {\relax{P.
  Tzallas}}(2023)}]{Tzallas_NatPhoton-NewsViews_2024}%
  \BibitemOpen
  \bibfield  {author} {\bibinfo {author} {\bibnamefont {\relax{P. Tzallas}}},\
  }\bibfield  {title} {\bibinfo {title} {Squeezed light effect},\ }\href
  {https://doi.org/10.1038/s41566-023-01218-9} {\bibfield  {journal} {\bibinfo
  {journal} {Nat. Photon.}\ }\textbf {\bibinfo {volume} {17}},\ \bibinfo
  {pages} {463–464} (\bibinfo {year} {2023})}\BibitemShut {NoStop}%
\bibitem [{\citenamefont {\relax{M. Lewenstein, P. Balcou, M. Y. Ivanov, A.
  L’Huillier and P. B. Corkum}}(1994)}]{Lewenstein_PRA-3step_1994}%
  \BibitemOpen
  \bibfield  {author} {\bibinfo {author} {\bibnamefont {\relax{M. Lewenstein,
  P. Balcou, M. Y. Ivanov, A. L’Huillier and P. B. Corkum}}},\ }\bibfield
  {title} {\bibinfo {title} {Theory of high-harmonic generation by lowfrequency
  laser fields},\ }\href {https://doi.org/10.1103/PhysRevA.49.2117} {\bibfield
  {journal} {\bibinfo  {journal} {Phys. Rev. A}\ }\textbf {\bibinfo {volume}
  {49}},\ \bibinfo {pages} {2117} (\bibinfo {year} {1994})}\BibitemShut
  {NoStop}%
\bibitem [{\citenamefont {\relax{P. Salières, A. L'Huillier, P. Antoine, M.
  Lewenstein}}(1999)}]{Salieres_AAMOP_1999}%
  \BibitemOpen
  \bibfield  {author} {\bibinfo {author} {\bibnamefont {\relax{P. Salières, A.
  L'Huillier, P. Antoine, M. Lewenstein}}},\ }\bibfield  {title} {\bibinfo
  {title} {Study of the spatial and temporal coherence of high-order
  harmonics},\ }\href {https://doi.org/doi.org/10.1016/S1049-250X(08)60219-0}
  {\bibfield  {journal} {\bibinfo  {journal} {Advances In Atomic, Molecular,
  and Optical Physics}\ }\textbf {\bibinfo {volume} {41}},\ \bibinfo {pages}
  {83} (\bibinfo {year} {1999})}\BibitemShut {NoStop}%
\bibitem [{\citenamefont {Constant}\ \emph {et~al.}(1999)\citenamefont
  {Constant}, \citenamefont {Garzella}, \citenamefont {Breger}, \citenamefont
  {M\'evel}, \citenamefont {Dorrer}, \citenamefont {Le~Blanc}, \citenamefont
  {Salin},\ and\ \citenamefont {Agostini}}]{Constant_PRL_1999}%
  \BibitemOpen
  \bibfield  {author} {\bibinfo {author} {\bibfnamefont {E.}~\bibnamefont
  {Constant}}, \bibinfo {author} {\bibfnamefont {D.}~\bibnamefont {Garzella}},
  \bibinfo {author} {\bibfnamefont {P.}~\bibnamefont {Breger}}, \bibinfo
  {author} {\bibfnamefont {E.}~\bibnamefont {M\'evel}}, \bibinfo {author}
  {\bibfnamefont {C.}~\bibnamefont {Dorrer}}, \bibinfo {author} {\bibfnamefont
  {C.}~\bibnamefont {Le~Blanc}}, \bibinfo {author} {\bibfnamefont
  {F.}~\bibnamefont {Salin}},\ and\ \bibinfo {author} {\bibfnamefont
  {P.}~\bibnamefont {Agostini}},\ }\bibfield  {title} {\bibinfo {title}
  {Optimizing high harmonic generation in absorbing gases: Model and
  experiment},\ }\href {https://doi.org/10.1103/PhysRevLett.82.1668} {\bibfield
   {journal} {\bibinfo  {journal} {Phys. Rev. Lett.}\ }\textbf {\bibinfo
  {volume} {82}},\ \bibinfo {pages} {1668} (\bibinfo {year}
  {1999})}\BibitemShut {NoStop}%
\bibitem [{\citenamefont {\relax{R. Weissenbilder, S. Carlström, L. Rego, et
  al.,}}(2022)}]{Huillier_NatRevPhys_2022}%
  \BibitemOpen
  \bibfield  {author} {\bibinfo {author} {\bibnamefont {\relax{R.
  Weissenbilder, S. Carlström, L. Rego, et al.,}}},\ }\bibfield  {title}
  {\bibinfo {title} {How to optimize high-order harmonic generation in gases},\
  }\href {https://doi.org/10.1038/s42254-022-00522-7} {\bibfield  {journal}
  {\bibinfo  {journal} {Nat. Rev. Phys.}\ }\textbf {\bibinfo {volume} {4}},\
  \bibinfo {pages} {713–722} (\bibinfo {year} {2022})}\BibitemShut {NoStop}%
\bibitem [{\citenamefont {\relax{I. Abram}}(1987)}]{Adam_PRA_1987}%
  \BibitemOpen
  \bibfield  {author} {\bibinfo {author} {\bibnamefont {\relax{I. Abram}}},\
  }\bibfield  {title} {\bibinfo {title} {Quantum theory of light propagation:
  Linear medium},\ }\href {https://doi.org/doi.org/10.1103/PhysRevA.35.4661}
  {\bibfield  {journal} {\bibinfo  {journal} {Phys. Rev. A}\ }\textbf {\bibinfo
  {volume} {35}},\ \bibinfo {pages} {4661} (\bibinfo {year}
  {1987})}\BibitemShut {NoStop}%
\bibitem [{\citenamefont {\relax{R. J. Glauber and M.
  Lewenstein}}(1991)}]{GlauberMaciej_PRA_1991}%
  \BibitemOpen
  \bibfield  {author} {\bibinfo {author} {\bibnamefont {\relax{R. J. Glauber
  and M. Lewenstein}}},\ }\bibfield  {title} {\bibinfo {title} {Quantum optics
  of dielectric media},\ }\href
  {https://doi.org/doi.org/10.1103/PhysRevA.43.467} {\bibfield  {journal}
  {\bibinfo  {journal} {Phys. Rev. A}\ }\textbf {\bibinfo {volume} {43}},\
  \bibinfo {pages} {467} (\bibinfo {year} {1991})}\BibitemShut {NoStop}%
\bibitem [{\citenamefont {\relax{J. Jeffers and S. M.
  Barnett}}(1994)}]{Jeffers_JMO_1994}%
  \BibitemOpen
  \bibfield  {author} {\bibinfo {author} {\bibnamefont {\relax{J. Jeffers and
  S. M. Barnett}}},\ }\bibfield  {title} {\bibinfo {title} {Propagation of
  squeezed light in dielectrics},\ }\href
  {https://doi.org/doi.org/10.1080/09500349414551061} {\bibfield  {journal}
  {\bibinfo  {journal} {J. Mod. Opt.}\ }\textbf {\bibinfo {volume} {41}},\
  \bibinfo {pages} {1121} (\bibinfo {year} {1994})}\BibitemShut {NoStop}%
\bibitem [{\citenamefont {\relax{M.T. Manzoni, D.E. Chang and J.S.
  Douglas}}(2017)}]{Manzoni_NatCommun_2017}%
  \BibitemOpen
  \bibfield  {author} {\bibinfo {author} {\bibnamefont {\relax{M.T. Manzoni,
  D.E. Chang and J.S. Douglas}}},\ }\bibfield  {title} {\bibinfo {title}
  {Simulating quantum light propagation through atomic ensembles using matrix
  product states},\ }\href {https://doi.org/doi.org/10.1038/s41467-017-01416-4}
  {\bibfield  {journal} {\bibinfo  {journal} {Nat. Commun}\ }\textbf {\bibinfo
  {volume} {8}},\ \bibinfo {pages} {1743} (\bibinfo {year} {2017})}\BibitemShut
  {NoStop}%
\bibitem [{\citenamefont {\relax{M. G. Raymer}}(2020)}]{Raymer_JMO_2020}%
  \BibitemOpen
  \bibfield  {author} {\bibinfo {author} {\bibnamefont {\relax{M. G.
  Raymer}}},\ }\bibfield  {title} {\bibinfo {title} {Quantum theory of light in
  a dispersive structured linear dielectric: a macroscopic hamiltonian tutorial
  treatment},\ }\href {https://doi.org/doi.org/10.1080/09500340.2019.1706773}
  {\bibfield  {journal} {\bibinfo  {journal} {J. Mod. Opt.}\ }\textbf {\bibinfo
  {volume} {67}},\ \bibinfo {pages} {196} (\bibinfo {year} {2020})}\BibitemShut
  {NoStop}%
\bibitem [{\citenamefont {Leonhardt}(1993)}]{leonhardt_quantum_1993}%
  \BibitemOpen
  \bibfield  {author} {\bibinfo {author} {\bibfnamefont {U.}~\bibnamefont
  {Leonhardt}},\ }\bibfield  {title} {\bibinfo {title} {Quantum statistics of a
  lossless beam splitter: {SU}(2) symmetry in phase space},\ }\href
  {https://doi.org/10.1103/PhysRevA.48.3265} {\bibfield  {journal} {\bibinfo
  {journal} {Physical Review A}\ }\textbf {\bibinfo {volume} {48}},\ \bibinfo
  {pages} {3265} (\bibinfo {year} {1993})}\BibitemShut {NoStop}%
\bibitem [{\citenamefont {Hergott}\ \emph {et~al.}(2002)\citenamefont
  {Hergott}, \citenamefont {Kovacev}, \citenamefont {Merdji}, \citenamefont
  {Hubert}, \citenamefont {Mairesse}, \citenamefont {Jean}, \citenamefont
  {Breger}, \citenamefont {Agostini}, \citenamefont {Carr\'e},\ and\
  \citenamefont {Sali\`eres}}]{Hergott_PRA_2002}%
  \BibitemOpen
  \bibfield  {author} {\bibinfo {author} {\bibfnamefont {J.-F.}\ \bibnamefont
  {Hergott}}, \bibinfo {author} {\bibfnamefont {M.}~\bibnamefont {Kovacev}},
  \bibinfo {author} {\bibfnamefont {H.}~\bibnamefont {Merdji}}, \bibinfo
  {author} {\bibfnamefont {C.}~\bibnamefont {Hubert}}, \bibinfo {author}
  {\bibfnamefont {Y.}~\bibnamefont {Mairesse}}, \bibinfo {author}
  {\bibfnamefont {E.}~\bibnamefont {Jean}}, \bibinfo {author} {\bibfnamefont
  {P.}~\bibnamefont {Breger}}, \bibinfo {author} {\bibfnamefont
  {P.}~\bibnamefont {Agostini}}, \bibinfo {author} {\bibfnamefont
  {B.}~\bibnamefont {Carr\'e}},\ and\ \bibinfo {author} {\bibfnamefont
  {P.}~\bibnamefont {Sali\`eres}},\ }\bibfield  {title} {\bibinfo {title}
  {Extreme-ultraviolet high-order harmonic pulses in the microjoule range},\
  }\href {https://doi.org/10.1103/PhysRevA.66.021801} {\bibfield  {journal}
  {\bibinfo  {journal} {Phys. Rev. A}\ }\textbf {\bibinfo {volume} {66}},\
  \bibinfo {pages} {021801} (\bibinfo {year} {2002})}\BibitemShut {NoStop}%
\bibitem [{\citenamefont {\relax{A. Zaïr, M. Holler, A. Guandalini, F.
  Schapper, J. Biegert†, L. Gallmann, U. Keller, A. S. Wyatt, A. Monmayrant,
  I. A. Walmsley, E. Cormier, T. Auguste, J. P. Caumes, and P.
  Salières}}(2008)}]{Zair_PRA_2008}%
  \BibitemOpen
  \bibfield  {author} {\bibinfo {author} {\bibnamefont {\relax{A. Zaïr, M.
  Holler, A. Guandalini, F. Schapper, J. Biegert†, L. Gallmann, U. Keller, A.
  S. Wyatt, A. Monmayrant, I. A. Walmsley, E. Cormier, T. Auguste, J. P.
  Caumes, and P. Salières}}},\ }\bibfield  {title} {\bibinfo {title} {Quantum
  path interferences in high-order harmonic generation},\ }\href
  {https://doi.org/doi.org/10.1103/PhysRevLett.100.143902} {\bibfield
  {journal} {\bibinfo  {journal} {Phys. Rev. Lett.}\ }\textbf {\bibinfo
  {volume} {100}},\ \bibinfo {pages} {143902} (\bibinfo {year}
  {2008})}\BibitemShut {NoStop}%
\bibitem [{\citenamefont {\relax{Anne L’Huillier and Philippe Balcou and
  Sebastien Candel and Kenneth J. Schafer and Kenneth C.
  Kulander}}(1992)}]{propag}%
  \BibitemOpen
  \bibfield  {author} {\bibinfo {author} {\bibnamefont {\relax{Anne
  L’Huillier and Philippe Balcou and Sebastien Candel and Kenneth J. Schafer
  and Kenneth C. Kulander}}},\ }\bibfield  {title} {\bibinfo {title}
  {\relax{Calculations of high-order harmonic-generation processes in xenon at
  1064 nm}},\ }\href {https://doi.org/10.1103/PhysRevA.46.2778} {\bibfield
  {journal} {\bibinfo  {journal} {Phys. Rev. A}\ }\textbf {\bibinfo {volume}
  {46}},\ \bibinfo {pages} {2778} (\bibinfo {year} {1992})}\BibitemShut
  {NoStop}%
\bibitem [{\citenamefont {\relax{Muffett, J. E. and Wahlström, Claes-Göran
  and Hutchinson, M. H. R}}(1994)}]{muffet}%
  \BibitemOpen
  \bibfield  {author} {\bibinfo {author} {\bibnamefont {\relax{Muffett, J. E.
  and Wahlström, Claes-Göran and Hutchinson, M. H. R}}},\ }\bibfield  {title}
  {\bibinfo {title} {\relax{Numerical Modeling of the Spatial Profiles of
  High-order Harmonics}},\ }\href {https://doi.org/10.1088/0953-4075/27/23/013}
  {\bibfield  {journal} {\bibinfo  {journal} {J. Phys. B: At. Mol. Opt. Phys.}\
  }\textbf {\bibinfo {volume} {27}},\ \bibinfo {pages} {5693} (\bibinfo {year}
  {1994})}\BibitemShut {NoStop}%
\bibitem [{\citenamefont {\relax{J. Peatross and D. D.
  Meyerhofer}}(1995)}]{peatheor}%
  \BibitemOpen
  \bibfield  {author} {\bibinfo {author} {\bibnamefont {\relax{J. Peatross and
  D. D. Meyerhofer}}},\ }\bibfield  {title} {\bibinfo {title}
  {\relax{Intensity-dependent atomic-phase effects in high-order harmonic
  generation}},\ }\href {https://doi.org/10.1103/PhysRevA.52.3976} {\bibfield
  {journal} {\bibinfo  {journal} {Phys. Rev. A}\ }\textbf {\bibinfo {volume}
  {52}},\ \bibinfo {pages} {3976} (\bibinfo {year} {1995})}\BibitemShut
  {NoStop}%
\bibitem [{\citenamefont {\relax{J. Peatross and M. V. Fedorov and K. C.
  Kulander}}(1995)}]{peatheor2}%
  \BibitemOpen
  \bibfield  {author} {\bibinfo {author} {\bibnamefont {\relax{J. Peatross and
  M. V. Fedorov and K. C. Kulander}}},\ }\bibfield  {title} {\bibinfo {title}
  {\relax{Intensity-dependent phase-matching effects in harmonic generation}},\
  }\href {https://doi.org/10.1364/JOSAB.12.000863} {\bibfield  {journal}
  {\bibinfo  {journal} {J. Opt. Soc. Am. B}\ }\textbf {\bibinfo {volume}
  {12}},\ \bibinfo {pages} {863} (\bibinfo {year} {1995})}\BibitemShut
  {NoStop}%
\bibitem [{\citenamefont {\relax{S. C. Rae, X. Chen, and K.
  Burnett}}(1994)}]{rae1}%
  \BibitemOpen
  \bibfield  {author} {\bibinfo {author} {\bibnamefont {\relax{S. C. Rae, X.
  Chen, and K. Burnett}}},\ }\bibfield  {title} {\bibinfo {title}
  {\relax{Saturation of harmonic generation in one- and three-dimensional
  atoms}},\ }\href {https://doi.org/10.1103/PhysRevA.50.1946} {\bibfield
  {journal} {\bibinfo  {journal} {Phys. Rev. A}\ }\textbf {\bibinfo {volume}
  {50}},\ \bibinfo {pages} {1946} (\bibinfo {year} {1994})}\BibitemShut
  {NoStop}%
\bibitem [{\citenamefont {\relax{S. C. Rae and K. Burnett and J.
  Cooper}}(1994)}]{rae}%
  \BibitemOpen
  \bibfield  {author} {\bibinfo {author} {\bibnamefont {\relax{S. C. Rae and K.
  Burnett and J. Cooper}}},\ }\bibfield  {title} {\bibinfo {title}
  {\relax{Generation and propagation of high-order harmonics in a rapidly
  ionizing medium}},\ }\href {https://doi.org/10.1103/PhysRevA.50.3438}
  {\bibfield  {journal} {\bibinfo  {journal} {Phys. Rev. A}\ }\textbf {\bibinfo
  {volume} {50}},\ \bibinfo {pages} {3438} (\bibinfo {year}
  {1994})}\BibitemShut {NoStop}%
\bibitem [{\citenamefont {{Maciej Lewenstein and Pascal Salières and Anne
  L'Huillier}}(1997)}]{phase}%
  \BibitemOpen
  \bibfield  {author} {\bibinfo {author} {\bibnamefont {{Maciej Lewenstein and
  Pascal Salières and Anne L'Huillier}}},\ }\bibfield  {title} {\bibinfo
  {title} {Phase of the atomic polarization in high-order harmonic
  generation},\ }\bibfield  {journal} {\bibinfo  {journal} {Phys. Rev. A}\
  }\textbf {\bibinfo {volume} {4747}},\ \href
  {https://doi.org/10.1103/PhysRevA.52.4747} {10.1103/PhysRevA.52.4747}
  (\bibinfo {year} {1997})\BibitemShut {NoStop}%
\bibitem [{\citenamefont {\relax{P. Salières and A. L'Huillier and M.
  Lewenstein}}(1995)}]{prlphase}%
  \BibitemOpen
  \bibfield  {author} {\bibinfo {author} {\bibnamefont {\relax{P. Salières and
  A. L'Huillier and M. Lewenstein}}},\ }\bibfield  {title} {\bibinfo {title}
  {\relax{Coherence Control of High-Order Harmonics}},\ }\href
  {https://doi.org/10.1103/PhysRevLett.74.3776} {\bibfield  {journal} {\bibinfo
   {journal} {Phys. Rev. Lett.}\ }\textbf {\bibinfo {volume} {75}},\ \bibinfo
  {pages} {3376} (\bibinfo {year} {1995})}\BibitemShut {NoStop}%
\bibitem [{\citenamefont {Carlström}\ \emph
  {et~al.}(2022{\natexlab{a}})\citenamefont {Carlström}, \citenamefont
  {Spanner},\ and\ \citenamefont {Patchkovskii}}]{carlstrom_general_2022}%
  \BibitemOpen
  \bibfield  {author} {\bibinfo {author} {\bibfnamefont {S.}~\bibnamefont
  {Carlström}}, \bibinfo {author} {\bibfnamefont {M.}~\bibnamefont
  {Spanner}},\ and\ \bibinfo {author} {\bibfnamefont {S.}~\bibnamefont
  {Patchkovskii}},\ }\bibfield  {title} {\bibinfo {title} {General
  time-dependent configuration-interaction singles. {I}. {Molecular} case},\
  }\href {https://doi.org/10.1103/PhysRevA.106.043104} {\bibfield  {journal}
  {\bibinfo  {journal} {Physical Review A}\ }\textbf {\bibinfo {volume}
  {106}},\ \bibinfo {pages} {043104} (\bibinfo {year}
  {2022}{\natexlab{a}})}\BibitemShut {NoStop}%
\bibitem [{\citenamefont {Carlström}\ \emph
  {et~al.}(2022{\natexlab{b}})\citenamefont {Carlström}, \citenamefont
  {Bertolino}, \citenamefont {Dahlström},\ and\ \citenamefont
  {Patchkovskii}}]{carlstrom_general_2022-1}%
  \BibitemOpen
  \bibfield  {author} {\bibinfo {author} {\bibfnamefont {S.}~\bibnamefont
  {Carlström}}, \bibinfo {author} {\bibfnamefont {M.}~\bibnamefont
  {Bertolino}}, \bibinfo {author} {\bibfnamefont {J.~M.}\ \bibnamefont
  {Dahlström}},\ and\ \bibinfo {author} {\bibfnamefont {S.}~\bibnamefont
  {Patchkovskii}},\ }\bibfield  {title} {\bibinfo {title} {General
  time-dependent configuration-interaction singles. {II}. {Atomic} case},\
  }\href {https://doi.org/10.1103/PhysRevA.106.042806} {\bibfield  {journal}
  {\bibinfo  {journal} {Physical Review A}\ }\textbf {\bibinfo {volume}
  {106}},\ \bibinfo {pages} {042806} (\bibinfo {year}
  {2022}{\natexlab{b}})}\BibitemShut {NoStop}%
\bibitem [{\citenamefont {Ilkov}\ \emph {et~al.}(1992)\citenamefont {Ilkov},
  \citenamefont {Decker},\ and\ \citenamefont {Chin}}]{ilkov_ionization_1992}%
  \BibitemOpen
  \bibfield  {author} {\bibinfo {author} {\bibfnamefont {F.~A.}\ \bibnamefont
  {Ilkov}}, \bibinfo {author} {\bibfnamefont {J.~E.}\ \bibnamefont {Decker}},\
  and\ \bibinfo {author} {\bibfnamefont {S.~L.}\ \bibnamefont {Chin}},\
  }\bibfield  {title} {\bibinfo {title} {Ionization of atoms in the tunnelling
  regime with experimental evidence using {Hg} atoms},\ }\href
  {https://doi.org/10.1088/0953-4075/25/19/011} {\bibfield  {journal} {\bibinfo
   {journal} {Journal of Physics B: Atomic, Molecular and Optical Physics}\
  }\textbf {\bibinfo {volume} {25}},\ \bibinfo {pages} {4005} (\bibinfo {year}
  {1992})}\BibitemShut {NoStop}%
\bibitem [{\citenamefont {Perelomov}\ \emph {et~al.}(1965)\citenamefont
  {Perelomov}, \citenamefont {Popov},\ and\ \citenamefont
  {Terent'ev}}]{perelomov_ionization_1967}%
  \BibitemOpen
  \bibfield  {author} {\bibinfo {author} {\bibfnamefont {A.}~\bibnamefont
  {Perelomov}}, \bibinfo {author} {\bibfnamefont {V.}~\bibnamefont {Popov}},\
  and\ \bibinfo {author} {\bibfnamefont {M.}~\bibnamefont {Terent'ev}},\
  }\bibfield  {title} {\bibinfo {title} {{Ionization of atoms in an alternating
  electric field: II}},\ }\href
  {http://www.jetp.ac.ru/cgi-bin/e/index/e/24/1/p207?a=list} {\bibfield
  {journal} {\bibinfo  {journal} {Sov. Phys. JETP}\ }\textbf {\bibinfo {volume}
  {24}},\ \bibinfo {pages} {207} (\bibinfo {year} {1965})}\BibitemShut
  {NoStop}%
\bibitem [{\citenamefont {Drummond}\ and\ \citenamefont
  {Gardiner}(1980)}]{drummond_generalised_1980}%
  \BibitemOpen
  \bibfield  {author} {\bibinfo {author} {\bibfnamefont {P.~D.}\ \bibnamefont
  {Drummond}}\ and\ \bibinfo {author} {\bibfnamefont {C.~W.}\ \bibnamefont
  {Gardiner}},\ }\bibfield  {title} {\bibinfo {title} {Generalised
  {P}-representations in quantum optics},\ }\href
  {https://doi.org/10.1088/0305-4470/13/7/018} {\bibfield  {journal} {\bibinfo
  {journal} {Journal of Physics A: Mathematical and General}\ }\textbf
  {\bibinfo {volume} {13}},\ \bibinfo {pages} {2353} (\bibinfo {year}
  {1980})}\BibitemShut {NoStop}%
\bibitem [{\citenamefont {Pisanty}(2020)}]{RBSFA}%
  \BibitemOpen
  \bibfield  {author} {\bibinfo {author} {\bibfnamefont {E.}~\bibnamefont
  {Pisanty}},\ }\href {https://doi.org/10.5281/zenodo.592519} {\bibinfo {title}
  {{RB-SFA: High Harmonic Generation in the Strong Field Approximation via
  Mathematica}}},\ \bibinfo {howpublished} {Github:
  \url{https://github.com/episanty/RB-SFA}} (\bibinfo {year}
  {2020})\BibitemShut {NoStop}%
\bibitem [{\citenamefont {\relax{B. Henke, E. Gullikson, and J.
  Davis}}(1993)}]{Hanke_IonData_1993}%
  \BibitemOpen
  \bibfield  {author} {\bibinfo {author} {\bibnamefont {\relax{B. Henke, E.
  Gullikson, and J. Davis}}},\ }\bibfield  {title} {\bibinfo {title} {X-ray
  interactions: photoabsorption, scattering, transmission, and reflection at e
  = 50–30,000 ev, z = 1–92},\ }\href@noop {} {\bibfield  {journal}
  {\bibinfo  {journal} {At. Data Nucl. Data Tables}\ }\textbf {\bibinfo
  {volume} {54}},\ \bibinfo {pages} {181} (\bibinfo {year} {1993})}\BibitemShut
  {NoStop}%
\end{thebibliography}%

\newpage
\clearpage
\onecolumngrid
\appendix

\begin{center}
    {\large \textbf{\textsc{Appendix}}}
\end{center}

\section{PROPAGATION FROM A SEMICLASSICAL PERSPECTIVE}
\subsection{Macroscopic response}
To calculate the macroscopic response of the system, one has to solve
the Maxwell equations for the fundamental and harmonic fields. This can be done using the slowly varying envelope and paraxial approximations. The fundamentals of such an approach have been formulated by \cite{propag}. Several groups have used similar approaches to study the effects of phase matching, and to perform direct comparison of the theory with experiments \cite{muffet,peatheor,peatheor2,rae1,rae}. One should stress that the theory presented in this paper, although based on semi-classical ideas, is very different, since it includes the quantum electrodynamical (QED) nature of the electromagnetic (EM) fields.

In the following, we present the Maxwell equations for the fundamental and harmonic fields used in the main text.~Using the slowly varying envelope and paraxial approximations, the propagation equations can be reduced to the form (here we use SI units)
\begin{align}\label{prop6}
&\nabla_\bot ^2 \boldsymbol{E}_1(\boldsymbol{r},t)
+ 2k_1^0 \delta k_1(\boldsymbol{r},t) 
\boldsymbol{E}_1(\boldsymbol{r},t)
+2ik_1^0{\partial \boldsymbol{E}_1(\boldsymbol{r},t) \over\partial z}= 0, 
\\&\nabla_\bot ^2 \boldsymbol{E}_q(\boldsymbol{r},t)
+  2 k_q^0 [\Delta k_q^0(z)+ \delta k_q(\boldsymbol{r},t)]\boldsymbol{E}_q(\boldsymbol{r},t)
+2i k_q^0{\partial \boldsymbol{E}_q(\boldsymbol{r},t) \over\partial z}
= - {q^2\omega^2\over \epsilon_0 c^2}\boldsymbol{P}_q(\boldsymbol{r},t),\label{prop7}
\end{align}
where $\boldsymbol{E}_1(\boldsymbol{r},t)$, and $\boldsymbol{E}_q(\boldsymbol{r} ,t)$ denote the slowly varying
(complex) envelopes of the fundamental and harmonic fields, respectively, $k_q^0=
q\omega/c$, whereas the rest of the symbols are explained below. The slow time dependence in the  above equations accounts for the temporal profile of the fundamental 
field that enters Eq.~\eqref{prop6} through the boundary condition
for $\boldsymbol{E}_1$. The solutions of the propagation equations for given $t$ have therefore
to be integrated over time.

The terms containing $ \Delta k_q^0(\boldsymbol{r},t)$ describe dispersion effects due to the linear polarisability of atoms, and can be in fact neglected in the
regime of parameters considered (low density). The terms proportional to $ \delta k_q(\boldsymbol{r},t)= -e^2{\cal N}_e(\boldsymbol{r},t) /(2mqc\omega)$, with $e$ denoting the  electron
charge, $m$- its   mass, and ${\cal N}_e(\boldsymbol{r},t)$ the electronic density,
describe the corrections to the index of refraction due to ionization; here the ionic part of those corrections is neglected. The electronic density
is equal to the number of ionized atoms 
\begin{equation}
{\cal N}_e(\boldsymbol{r},t)= {\cal N}_a(z)\left[1-\exp\left(-\int_{-\infty}^t
\Gamma(|\boldsymbol{E}_1(\boldsymbol{r},t')|)\,\dd t'\right)\right],
\end{equation}
where ${\cal N}_a(z)$ is the initial density of the atomic jet and $\Gamma(|\boldsymbol{E}_1(\boldsymbol{r},t')|)$ is the total ionization rate, taking into account the contributions of all active electrons using $\Gamma=2 {\rm Re}[\int_0^T\gamma(t)\dd t/T]
$ with $T=2\pi/\omega$ for an instantaneous and local value of the electric field envelope  $\boldsymbol{E}_1(\boldsymbol{r},t')$.~Note that since $\Gamma$ depends functionally on $\boldsymbol{E}_1(\boldsymbol{r},t')$, Eq.~\eqref{prop6} is a nonlinear integro-differential equation; it has to be solved first, and its solution is used then to solve Eq.~\eqref{prop7}.

Finally, the Fourier components of the atomic polarization are given
by
\begin{equation}\label{polar}
    \begin{aligned}
&\boldsymbol{P}_q(\boldsymbol{r},t)=
2{\cal N}_a(z) \boldsymbol{x}_q(\boldsymbol{r},t)e^{iq\phi_1(\boldsymbol{r}, t) }\exp\left(-\int_{-\infty}^t
\Gamma(|\boldsymbol{E}_1(\boldsymbol{r},t')|)\,\dd t'\right),\nonumber
\end{aligned}
\end{equation}
where $\boldsymbol{x}_q(\boldsymbol{r},t)$ denotes the harmonic components of the total atomic dipole moment, which includes the couplings of all active electrons calculated for a field
($|E_{1x}| \cos(\omega t), |E_{1y}|\sin(\omega t), 0$).~The factor of 2 arises from the different conventions used in the definitions of $\boldsymbol{P}_q$ and $\boldsymbol{x}_q$.~Finally, $\phi_1(\boldsymbol{r},t)$ represents the phase of the laser field envelope $\boldsymbol{E}_1 (\boldsymbol{r},t)$, obtained by solving the propagation equation for the fundamental.

\subsection{Phase matching effects}
Phase matching effects depend on many aspects of the HHG process, but in the first place on the dynamically induced phase of the atomic polarization~\cite{phase,prlphase}. This phase, in general, is ``random'', since it includes interference of various electronic trajectories. This can be influenced by the propagation effects, leading to phase matching and generation of attosecond pulse trains. It can also be compensated by the intrinsic Guoy phase of the driving laser pulse \cite{prlphase}, leading to on-axis or off-axis generation, for harmonics corresponding to different families of electronic trajectories.   Propagation might decrease the extent of the plateau of the harmonic spectrum compared to the single atom response, from a photon energy of $I_p+3.2U_p$ to about $I_p+2U_p$.

\section{QUANTUM OPTICAL ANALYSIS}
\subsection{Husimi distribution for BSV light}\label{Sec:App:Husimi:function}
The Husimi distribution as a function of the electric field amplitude reads,
\begin{equation}\label{Eq:BSVHusimi}
	\begin{split}
		Q(E)= \frac{1}{\pi\cosh(r)}\exp[-\frac{ E_{x_{1}}^2}{2(1+e^{2r})}-\frac{E_{x_{2}}^2}{2(1+e^{-2r})}],
	\end{split}
\end{equation}
where $r$ is the squeezing parameter. The $Q(E)$ is a marginal of the Husimi function $\pi^{-1}\mel{\alpha}{\hat{\rho}}{\alpha}$, with $\hat{\rho}$. The sign of $r$ (whether positive or negative) determines which quadrature is squeezed, namely $x_1$ or $x_2$. However, for BSV states---being centered at the origin of phase space and lacking a coherent displacement---there is no intrinsic phase reference. Consequently, the distinction between ``amplitude'' and ``phase'' squeezing is arbitrary, and all orientations of squeezing are virtually equivalent. Thus, without loss of generalization, we set $r>0$. Furthermore, besides having a vanishing mean field amplitude, the mean photon number of the state is $\langle n \rangle=\text{sinh}^2(r)$, related to the mean intensity through $\langle I\rangle=c\hbar\omega\langle n \rangle/V$, where $V$ is the quantization volume, considered here to be about $10^{-14} $ cm$^3$.

\subsection{Ionization rates}\label{Sec:App:Ion:rates}

\subsubsection{Ionization rates for classical fields}
Here, we work in the quasistatic approximation, assuming that the ionization rate $\Gamma$ depends only on the field strength, and parametrically on the frequency $\omega$, denoted as $\Gamma[E(t)]$. For an atom initially in its ground state, the ionization yield can be obtained as
\begin{equation}\label{Eq:Methods:Yi}
    \pdv{Y_i(t)}{t}
        = \Gamma[E(t)]
            \left[
                1-Y_i(t)
            \right], \quad Y_i(-\infty) = 0,
\end{equation}
with solution
\begin{equation}
    Y_i(t)
        = 1 - \exp\left\{-\int^{t}_{-\infty} \dd \tau\ \Gamma[E(\tau)]\right\}.
\end{equation}

The calculations are performed within the time-dependent configuration interaction singles (TD-CIS) ansatz (see Refs.~\cite{carlstrom_general_2022,carlstrom_general_2022-1} for details). More specifically, if $\ket{\psi(t)}$ denotes the solution of the time-dependent Schrödinger equation (TDSE) at time $t$, we approximate the ionization yield as the norm lost from the calculation box, i.e., $Y_i(t) \approx 1- \abs{\psi(t+t_D)}^2$, where an additional time delay $t_D$ accounts for the finite time the ionized electron takes to leave the box. Consequently, from Eq.~\eqref{Eq:Methods:Yi}, we have
\begin{equation}
    \int^t_{-\infty} 
        \dd\tau\ \Gamma[E(t)]
            \approx - \ln[\abs{\psi(t+t_D)}^2].
\end{equation}
Neglecting sub-cycle effects, the integral above can be approximated as
\begin{equation}
    \int^t_{-\infty}
        \dd\tau\ \Gamma[E(t)]
            \approx f(t_0) + \Gamma(E_0)(t-t_0),
\end{equation}
and a straight-line fit to $-\ln[\abs{\psi(t+t_D)}^2]$ is used to extract the ionization rate $\Gamma(E_0)$. For numerical reasons, this approach yields TDSE ionization rates which are reliable above approximately $I_0 = 10^{13}$ W/cm$^2$. Below this intensity, we instead use PPT rates~\cite{ilkov_ionization_1992,perelomov_ionization_1967}, scaled to match the TDSE rates in the intermediate regime.

\subsubsection{Ionization probability under the influence of quantum light}
If $P_{\text{bound}}(t_0)$ denotes the probability of the system being in a bound state at the initial time $t_0$, then at any later time this probability satisfies
\begin{equation}
	\dv{P_{\text{bound}}(t)}{t}
		= - \Gamma(t) P_{\text{bound}}(t)
	\Rightarrow P_{\text{bound}}(t) 
		= \exp[-\int^t_{t_0}\dd \tau \Gamma(\tau)]
			P_{\text{bound}(t_0)},
\end{equation}
which is trivially related to the ionization probability via 
\begin{equation}
	P_{\text{ion}}(t) = 1 - P_{\text{bound}}(t).
\end{equation}
For a generic quantum state $\hat{\rho}(t)$ describing the initial state of the system at time $t$, this probability can equivalently be expressed as
\begin{equation}\label{Eq:Prob:ionization}
		P_{\text{ion}}(t) 
			= 1 - \tr(\sum_{i \in \text{bound}} \dyad{\psi_i} \hat{\rho}),
\end{equation}
where $\{\ket{\psi_i}: i\in \text{bound}\}$ denotes the set of all bound states of the atomic system.

When strong-field phenomena are driven with a generic state of light, it can be shown~\cite{Gorlach_NatPhys-HHG-BSV_2023,Javier_PRL-BSV_2025} that the final quantum state of the joint light-matter system can be written in terms of the generalized positive $P$ representation~\cite{drummond_generalised_1980} as
\begin{equation}
	\hat{\rho}(t)
		\approx
			\int \dd^2 \alpha \int \dd^2 \beta
				\dfrac{P(\alpha,\beta^*)}{\braket{\beta^*}{\alpha}}
					\dyad{\phi_{\alpha}(t)}{\phi_{\beta^*}(t)}
						\otimes \hat{\rho}_{\text{light}}(t),					
\end{equation}
with $\hat{\rho}_{\text{light}}(t)$ representing the quantum optical state, and $\ket{\phi_{\alpha}(t)}$ the electronic state evolving under the Hamiltonian $\hat{H} = \hat{H}_{\text{atom}} + \hat{\boldsymbol{d}}\cdot \mel{\alpha,\{0\}}{\hat{\boldsymbol{E}}(t)}{\alpha,\{0\}}$ where $\hat{E}(t)$ the electric field operator. Inserting this expression into Eq.~\eqref{Eq:Prob:ionization}, we obtain
\begin{equation}
	P_{\text{ion}}(t) 
		\approx 1 -
			\sum_{i\in \text{bound}}
				 \int \dd^2 \alpha \int \dd^2 \beta
					\dfrac{P(\alpha,\beta^*)}{\braket{\beta^*}{\alpha}}
						\langle\phi_{\beta^*}(t)\vert\psi_i\rangle\braket{\psi_i}{\phi_{\alpha}(t)}.
\end{equation}

Because we are dealing with strong-fields propagating in free space, we evaluate this expression in the classical limit~\cite{Gorlach_NatPhys-HHG-BSV_2023,Javier_PRL-BSV_2025}. This requires taking $V\to \infty$ and $\alpha\to\infty$, with $V$ denoting the quantization volume, while ensuring that the electric field strength $E_{\alpha} \propto \abs{\alpha}/\sqrt{V}$ remains finite. In this limit, we arrive at
\begin{equation}
	\begin{aligned}
	P_{\text{ion}}(t) 
		&\approx 1 -
			\sum_{i\in \text{bound}}
				\int \dd^2 E_\alpha 
					\Big[
						\lim_{V\to\infty}P(E_\alpha,E_\alpha)
					\Big]
					\abs{\langle\phi_{\alpha}(t)\vert\psi_i\rangle}^2
		\\&\approx 1 
			- \int \dd^2 E_\alpha 
						\Big[
							\lim_{V\to\infty}P(E_\alpha,E_\alpha)
						\Big]
						e^{-\int^{t}_{-\infty} \dd \tau \Gamma_{\text{ADK}}^{(\alpha)}(\tau)},
	\end{aligned}
\end{equation}
which we refer to as $\langle Y_i(t)\rangle$ in the main text.

\subsection{Analysis of the HHG spectrum}\label{Sec:App:HHG:Spec}
\subsubsection{Introducing depletion in the analysis of the HHG spectrum}
When driving HHG with very large intensities---typically exceeding $10^{14}$ W/cm$^2$---ground-state depletion effects become non-negligible and can significantly reduce the intensity of the emitted harmonics. Under such conditions, the standard SFA, which assumes negligible depletion, must be revised. In particular, the usual ansatz
\begin{equation}
	\ket{\psi(t)}
		= a(t) \ket{\text{g}}
			+\int \dd v \ b(v,t)\ket{v},
\end{equation}
requires a more accurate treatment of the ground-state amplitude $a(t)$. While typical SFA analyses assume $\abs{a(t)} \approx 1$, this approximation breaks down at high intensities. To account for depletion, we find that the ground-state amplitude evolves according to
\begin{equation}
	a(t) \propto e^{-\frac12\int^{t}_{-\infty} \dd \tau \Gamma(\tau)},
\end{equation}
up to a phase prefactor, where $\Gamma(t)$ denotes the ionization rate.

Consequently, when depletion effects are included, the time-dependent dipole moment---used to compute the HHG spectrum---takes the form
\begin{equation}\label{Eq:HHG:dipole}
	d(t)
		= \bra{\psi(t)}\hat{d}\ket{\psi(t)}
		\approx 
				\int \dd v \int_{t_0}^t \dd t_1
					e^{-\frac12\int^{t}_{-\infty} \dd \tau \Gamma(\tau)}
					e^{-iS(p,t,t_1)}
					d\big(p + A(t)\big)
					E(t_1)
					d\big(p + A(t_1)\big)
					e^{-\frac12\int^{t_1}_{-\infty} \dd \tau \Gamma(\tau)},
\end{equation}
with $S(p,t,t_1)$ is the semiclassical action, $E(t)$ the electric field, and $d(v)$ denotes the transition matrix element between the ground state and a continuum state $\ket{v}$. In our case, we compute the dipole using the \texttt{RB-SFA} Mathematica package~\cite{RBSFA}, chosen for its computational efficiency in generating HHG spectra under the SFA.

However, this efficiency relies on the saddle-point approximation, which does not account for depletion effects. To incorporate depletion, we exploit one of the package’s features---its ability to treat the vector potential in the semiclassical action and the electric field in the dipole matrix element as independent quantities. This flexibility allows us to define a modified electric field that incorporates depletion
\begin{equation}
	\Tilde{E}(t)
		= E(t)e^{-\frac12\int^{t}_{-\infty} \dd \tau \Gamma(\tau)},
\end{equation}
so that the dipole moment computed within \texttt{RB-SFA} becomes
\begin{equation}
	d_{\texttt{RB-SFA}}(t)
		\approx 
			\int \dd v \int_{t_0}^t \dd t_1
				e^{-iS(p,t,t_1)}
				d\big(p + A(t)\big)
				\Tilde{E}(t_1)
				d\big(p + A(t_1)\big),
\end{equation}
which relates to the full semiclassical dipole via $d(t) = e^{-\frac12\int^{t}_{-\infty} \dd \tau \Gamma(\tau)}d_{\texttt{RB-SFA}}(t)$. Thus, to fully account for depletion effects, we simply multiply the \texttt{RB-SFA} dipole by the ionization factor $e^{-\frac12\int^{t}_{-\infty} \dd \tau \Gamma(\tau)}$. The HHG spectrum is then obtained via a Fast Fourier Transform (FFT) of $d(t)$. Figure S1 shows the limiting $Q(\alpha) = \lim_{V\to\infty} P(\alpha,\alpha)$ function for different squeezing intensities (left $y$ axis) alongside the corresponding harmonic yield associated with the range of electric field strengths spanned by the Wigner function for various harmonics (right $y$ axis). As observed, a broader $Q(\alpha)$ distribution implies that a larger range of electric field strengths contributes; however, beyond a critical field strength, the harmonic yield decreases sharply.

\begin{figure}[h!]
	\centering
	\includegraphics[width=1\textwidth]{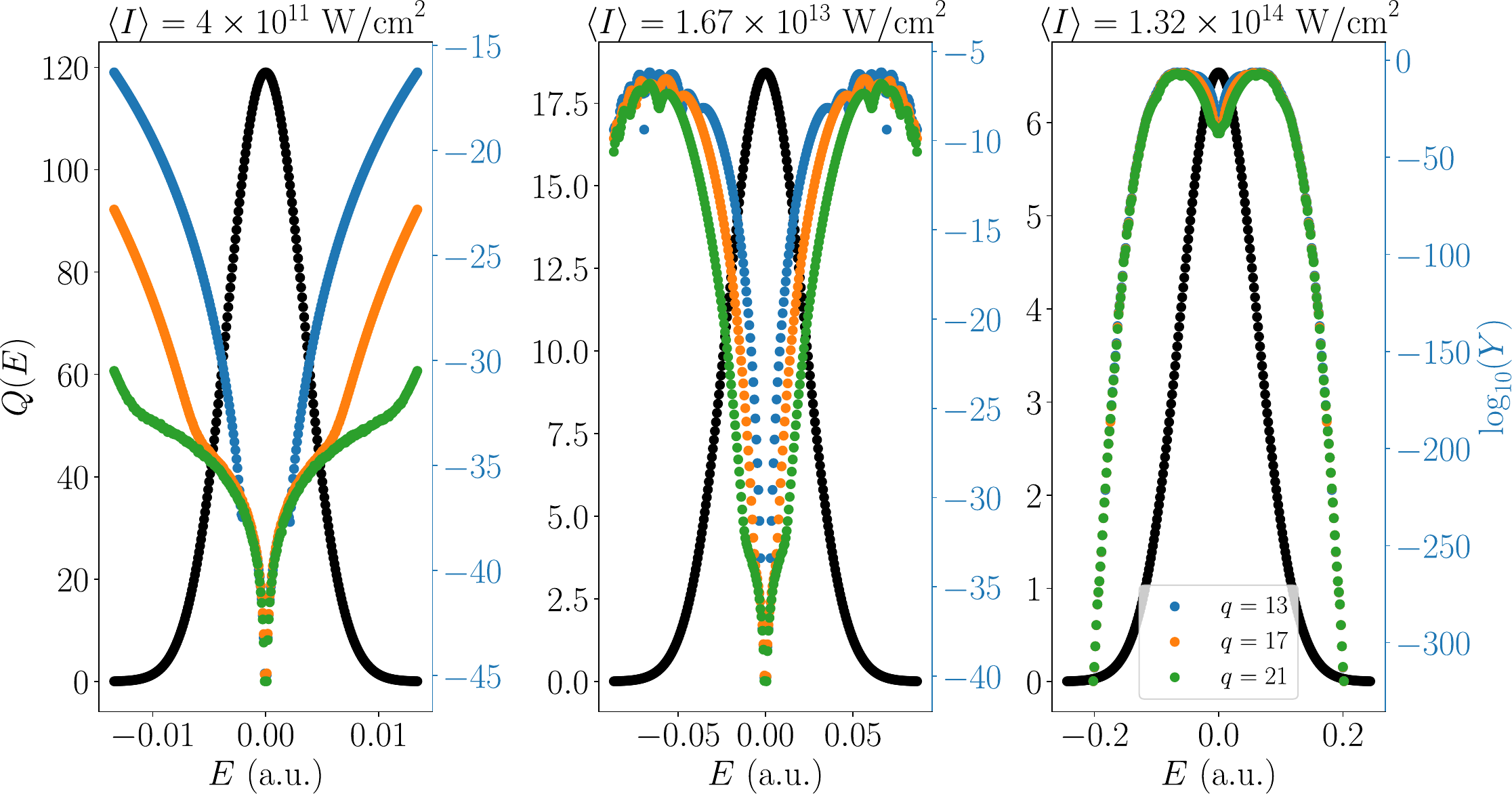}
	\caption{$Q(\alpha)$ function and harmonic yields of various harmonics computed for different mean intensities of the driving BSV light. In all cases, we considered a pulse with a $\sin^2$-envelope, 13 fs of duration and central wavelength $\lambda = 800$ nm. The atomic species under consideration is Argon ($I_p=0.58$ a.u.).}
	\label{Fig:HHG:Spec}
\end{figure}

\subsubsection{Propagation of the electric field}
The main objective of this section is to derive the following classical propagation equation for the $q$th harmonic order
\begin{equation}\label{Eq:paraxial:cl:prop}
	\laplacian_{\perp} E_q
		-2 i k_q \pdv{E_q}{z}
			= -\mu_0 q^2 \omega^2 P_q e^{i(qk_1 - k_q)z},
\end{equation}
where $z$ is the propagation direction, and $E_q$ and $P_q$ are the field amplitude and polarization components related to the $q$th harmonic order. First, we will try to understand where this equation comes from a classical perspective. Then, we will attempt to do a sort of quantum optical justification for using it in our analysis.\\

\noindent\textbf{Classical analysis}\\
For the classical derivation, our starting point are Maxwell equations in a source-free, non-magnetic media that, however, has a non-zero electrical susceptibility. These equations are
\begin{equation}
	\div{\boldsymbol{D}} = \boldsymbol{0},\quad
	\div{\boldsymbol{B}} = \boldsymbol{0},\quad
	\curl{\boldsymbol{B}} = \mu_0 \pdv{\boldsymbol{D}}{t},\quad
	\curl{\boldsymbol{E}} = - \pdv{\boldsymbol{B}}{t},
\end{equation} 
where $\boldsymbol{D} = \varepsilon_0 \boldsymbol{E} + \boldsymbol{P}$, where $\boldsymbol{P}$ represents the polarization. Combining these equations, we arrive at
\begin{equation}
	\curl{\curl{\boldsymbol{E}}}
		= -\mu_0 \varepsilon_0
			\pdv[2]{\boldsymbol{E}}{t}
				- \mu_0 \pdv[2]{\boldsymbol{P}}{t},
\end{equation}
which by using the relation
\begin{equation}
	\curl{\curl{\boldsymbol{E}}}
		= \grad\big(\div{\boldsymbol{E}}\big)
			- \laplacian{\boldsymbol{E}},
\end{equation}
can be rewritten as
\begin{equation}\label{Eq:WE:total}
	\laplacian{\boldsymbol{E}}
		-\grad\big(\div{\boldsymbol{E}}\big)
			=  \mu_0 \varepsilon_0
				\pdv[2]{\boldsymbol{E}}{t}
				+\mu_0 \pdv[2]{\boldsymbol{P}}{t}.
\end{equation}

Taking into account that $\boldsymbol{P} \propto \chi^{(n)} \boldsymbol{E}^n_L$, then from Gauss law we have that
\begin{equation}
	\div{\boldsymbol{D}}
		= \boldsymbol{0} \implies
	\div{\boldsymbol{E}}
		= \boldsymbol{0},
\end{equation}
such that Eq.~\eqref{Eq:WE:total} reads
\begin{equation}\label{Eq:Prop:cl}
	\laplacian{\boldsymbol{E}}
		=  \mu_0 \varepsilon_0
				\pdv[2]{\boldsymbol{E}}{t}
			+\mu_0 \pdv[2]{\boldsymbol{P}}{t}.
\end{equation}

This equation is, under the conditions established at the beginning, general and exact; no approximations have been done thus far. However, to arrive to Eq.~\eqref{Eq:paraxial:cl:prop} it is crucial to introduce the paraxial approximation, which essentially consists in neglecting second-order spatial derivatives along the propagation direction, thus focusing on what happens close to the propagation axis. When expanding the electric field in terms of the harmonic orders
\begin{equation}
	\boldsymbol{E}
		= \sum_q \boldsymbol{E}_q e^{i(k_q z + q\omega t)} + \text{c.c.},
\end{equation}
this results in
\begin{equation}
	\laplacian\big[\boldsymbol{E}_q e^{i k_q z}\big]
		\approx e^{i k_q z}
			\Big[
				\laplacian_\perp \boldsymbol{E}_q
					+ 2i k_q \pdv{\boldsymbol{E}_q}{z}
						-k_q^2 \boldsymbol{E}_q
			\Big],
\end{equation}
such that the propagation equation reads
\begin{equation}
	\laplacian_\perp{\boldsymbol{E}_q}
		+ 2i k_q \pdv{\boldsymbol{E}_q}{z}
			= \mu_0 \pdv[2]{\boldsymbol{P}}{t}.
\end{equation}

To conclude, we expand the polarization as
\begin{equation}
	\boldsymbol{P}
		= \sum_q
				\boldsymbol{P}_q
					e^{i(qk_L z + q \omega t)} + \text{c.c.},
\end{equation}
given that the $q$th harmonic order is affected by the $q$th susceptibility order of the material, proportional to $\boldsymbol{E}_L^q$. With this, we arrive at
\begin{equation}
	\laplacian_\perp{\boldsymbol{E}_q}
		+ 2i k_q \pdv{\boldsymbol{E}_q}{z}
			= - \mu_0(q\omega)^2 \boldsymbol{P}_q
					e^{i(qk_1-k_q)z},
\end{equation}
corresponding to Eq.~\eqref{Eq:paraxial:cl:prop} given initially.\\

\noindent\textbf{Quantum optical analysis}\\
To present a potential quantum optical formalism that we can make compatible with the propagation calculations we have considered thus far, we could consider as starting point the following light-matter interaction Hamiltonian
\begin{equation}
	\hat{H} 
		= \hat{H}_{\text{system}}
			+ \int \dd \vb{r}
					\hat{\boldsymbol{J}}(\vb{r}) \cdot \boldsymbol{A}(\vb{r},t)
			+ \dfrac12\int \dd \vb{r}
					\bigg\{
						\dfrac{1}{\epsilon_0} \hat{\boldsymbol{\Pi}}(\vb{r}) 
						+ \dfrac{1}{\mu_0}
							\big[
								\curl{\boldsymbol{A}(\vb{r})}
							\big]^2
					\bigg\}
			+ \hat{H}_{\text{other}},
\end{equation}
where the first term represents the system Hamiltonian, the second is the interaction term with $\boldsymbol{J}(\vb{r})$ the current operator, the third is the free-field Hamiltonian while the fourth contains additional terms such as the diamagnetic term. In the free-field Hamiltonian, $\hat{\boldsymbol{A}}(\vb{r})$ is the vector potential operator and $\hat{\boldsymbol{\Pi}}(\vb{r})$ is its conjugate variable. These two satisfy
\begin{equation}
	[\hat{A}_m(\vb{r}), \hat{\Pi}_n(\vb{r})]
		= i\hbar \delta_{m,n}^{\perp}(\vb{r}-\vb{r}'),
\end{equation}
with $ \delta_{m,n}^{\perp}(\vb{r}-\vb{r}')$ the transverse Dirac-delta function.

For our purposes, we want to look at how the $\hat{\boldsymbol{\Pi}}(\vb{r})$ operator, which in the end is related to the electric field operator, evolves in time. For that, we use the Heisenberg equation
\begin{equation}
	\partial_t \hat{\Pi}_i(\vb{r})
		= \dfrac{i}{\hbar} [\hat{H},\hat{\Pi}_i(\vb{r})].
\end{equation}
In the following, we evaluate the commutator with each of the elements of the Hamiltonian separately. Firstly, we have that $[\hat{H}_{\text{system}}, \hat{\Pi}_i(\vb{r})] = 0$, so we begin a more in-depth analysis for the free-field Hamiltonian. Considering hat part of this Hamiltonian commutes with $\hat{\Pi}_i(\vb{r})$, we find that
\begin{equation}\label{Eq:comn:freefield}
	[\hat{H}_{\text{field}}, \hat{\Pi}_i(\vb{r})]
		= \dfrac12
			\int \dd \vb{r}'
				\big[ 
					\big(\curl{\hat{\boldsymbol{A}}}(\vb{r}')\big)^2,\hat{\Pi}_i(\vb{r})
				\big].
\end{equation}

Having in mind that $(\curl{\hat{A}(\vb{r})})_j = \sum_{k,l} \varepsilon_{jkl} \partial_k \hat{A}_l(\vb{r})$ with $\varepsilon_{jkl}$ the Levi-Civita symbol, we can write
\begin{equation}
	\big(\curl{\hat{\boldsymbol{A}}}(\vb{r}')\big)^2
		= \sum_{jklmn}
				\varepsilon_{jkl}
					\varepsilon_{jmn}
						\partial'_k \hat{A}_l(\vb{r})
							\partial'_m \hat{A}_n(\vb{r}),
\end{equation}
such that we can write the commutator within the integral as
\begin{equation}
	\begin{aligned}
		[\big(\curl{\hat{\boldsymbol{A}}}(\vb{r}')\big)^2,\hat{\Pi}_i(\vb{r})]
			&= \sum_{jklmn}
					\varepsilon_{jkl}	
						\varepsilon_{jmn}
							\Big\{
								\partial'_k \hat{A}_l(\vb{r'})
									[\partial'_m \hat{A}_n(\vb{r}'),\hat{\Pi}_i(\vb{r})]
								+ [\partial'_k \hat{A}_l(\vb{r'}),\hat{\Pi}_i(\vb{r})]
										\partial'_m \hat{A}_n(\vb{r}')
							\Big\}
			\\& =i\hbar \sum_{jklmn}
					\varepsilon_{jkl}	
						\varepsilon_{jmn}
							\Big\{
								\partial'_k \hat{A}_l(\vb{r'})
								\partial'_m \delta_{n,i}^\perp(\vb{r}'-\vb{r})
								+ \partial'_k \delta_{l,i}^\perp(\vb{r}'-\vb{r})
								\partial'_m \hat{A}_n(\vb{r}')
							\Big\},
	\end{aligned}
\end{equation}
and after integrating by parts, we find for the integral
\begin{equation}
	\int \dd \vb{r}'
		[\big(\curl{\hat{\boldsymbol{A}}}(\vb{r}')\big)^2,\hat{\Pi}_i(\vb{r})]
			=-i\hbar \sum_{jklmn}
					\varepsilon_{jkl}	
						\varepsilon_{jmn}
							\int \dd \vb{r}'
								\bigg[
									\delta_{n,i}^\perp(\vb{r}'-\vb{r})
										\partial'_m \partial'_k
											\hat{A}_l(\vb{r}')
									+ \delta_{l,i}^\perp(\vb{r}'-\vb{r})
										\partial'_k\partial'_m \hat{A}_n(\vb{r}')
								\bigg].
\end{equation}

Now, having in mind the following property of the Levi-Civita symbols
\begin{equation}
	\sum_{j}\varepsilon_{jkl}\varepsilon_{jmn}
		= \delta_{km}\delta_{ln} - \delta_{kn}\delta_{lm},
\end{equation}
which introduced in our expressions results in
\begin{equation}
	\int \dd \vb{r}'
	[\big(\curl{\hat{\boldsymbol{A}}}(\vb{r}')\big)^2,\hat{\Pi}_i(\vb{r})]
		=-2i\hbar\sum_{m}
			\int \dd \vb{r}'
				\bigg[
					\delta^\perp(\vb{r}'-\vb{r})
						\partial'^2_m
								\hat{A}_l(\vb{r}')
					- \delta^\perp(\vb{r}'-\vb{r})
						\partial'_m\partial'_i \hat{A}_m(\vb{r}')
							\bigg],
\end{equation}
and that, after taking into account that $\partial_m\partial_n \hat{A}_i =\partial_n\partial_m \hat{A}_i$, we can more compactly write as
\begin{equation}
	\int \dd \vb{r}'
		[\big(\curl{\hat{\boldsymbol{A}}}(\vb{r}')\big)^2,\hat{\Pi}_i(\vb{r})]
			=-2i\hbar
				\int \dd \vb{r}' \delta^\perp(\vb{r}'-\vb{r})
					\big[
						\laplacian{\hat{A}_i(\vb{r}')}
						- \partial'_i \div{\hat{\boldsymbol{A}}(\vb{r}')}
					\big].
\end{equation}

Working in the Coulomb gauge implies that $\div{\hat{\boldsymbol{A}}(\vb{r}')} = 0$, such that the vector potential operator is a transverse operator. Consequently, the Dirac delta acts trivially, and we can write
\begin{equation}
	\int \dd \vb{r}'
		[\big(\curl{\hat{\boldsymbol{A}}}(\vb{r}')\big)^2,\hat{\Pi}_i(\vb{r})]
			= -2i\hbar \laplacian{\hat{A}_i(\vb{r})}.
\end{equation}
resulting in the following evolution of $\hat{\Pi}(\vb{r})$ in the absence of any other contribution
\begin{equation}
	\partial_{t} \hat{\Pi}_i(\vb{r})
		= \dfrac{1}{\mu_0}
			\laplacian{\hat{A}}_i(\vb{r}),
\end{equation}
and provided that $\hat{\Pi}_i(\vb{r}) = \varepsilon_0 \partial_t \hat{A}_i(\vb{r})$, if we derive the whole equation with respect to time we get
\begin{equation}
	\partial^2_{t} \hat{\Pi}_i(\vb{r})
	= \dfrac{1}{\mu_0\epsilon_0}
	\laplacian{\hat{\Pi}}_i(\vb{r}),
\end{equation}
which is the usual wave-equation.

The introduction of additional terms results in extra sources. For example, when incorporating the charge current we find
\begin{equation}
	\int \dd \vb{r}
		[\hat{\boldsymbol{J}}(\vb{r}) \cdot \boldsymbol{A}(\vb{r},t), \hat{\Pi}_i(\vb{r})]
			= i\hbar\int \dd \vb{r}' \hat{J}_i(\vb{r}) \delta^{\perp}(\vb{r}'-\vb{r})
			= i\hbar\hat{J}_i(\vb{r}), \quad \text{with}\ i \in \perp,
\end{equation}
where only the transverse contributions of the current should contribute as the vector potential is a transverse operator by definition. Consequently, the additional terms result in 
\begin{equation}
	\partial^2_{t} \hat{\Pi}_i(\vb{r})
		= \dfrac{1}{\mu_0\epsilon_0}
			\laplacian{\hat{\Pi}}_i(\vb{r})
			- \partial_t \hat{\boldsymbol{J}}_i(\vb{r})
			+ \dfrac{i}{\hbar}
					\partial_t[\hat{H}_{\text{others}},\hat{\Pi}_i(\vb{r})],
\end{equation}
after performing the additional derivative over time. In fact, having in mind that current and polarization are related through $\hat{\boldsymbol{J}}(\vb{r}) = \partial_t \hat{\boldsymbol{P}}(\vb{r})$, we can rewrite the equation as
\begin{equation}\label{Eq:App:evolv:Pi:op}
	\partial^2_{t} \hat{\Pi}_i(\vb{r})
		= \dfrac{1}{\mu_0\epsilon_0}
			\laplacian{\hat{\Pi}}_i(\vb{r})
				- \partial^2_t \hat{\boldsymbol{P}}_i(\vb{r})
					+ \dfrac{i}{\hbar}
	\partial_t[\hat{H}_{\text{others}},\hat{\Pi}_i(\vb{r})],
\end{equation}
which essentially resembles Eq.~\eqref{Eq:Prop:cl} when
\begin{itemize}
	\item Applying the Ehrenfest theorem such that we essentially deal with mean values;
	\item When neglecting the contribution of the commutator with $\hat{H}_{\text{others}}$.
\end{itemize}

\noindent\textbf{Propagation for the generated harmonics}\\
We are interested in the intensity of the harmonics at a given position $\vb{r}$ and time $t$.~Denoting by $\hat{\rho}_q$ the state of the $q$th harmonic order before propagating to the space-time point $(\vb{r},t)$, this quantity can be computed as
\begin{equation}
	\langle I(\vb{r},t)\rangle
		= \tr[\hat \rho \hat{\Pi}^2_{i,q}(\vb{r},t)],
\end{equation}
with $\hat{\Pi}_{i,q}(\vb{r},t)$ satisfying Eq.~\eqref{Eq:App:evolv:Pi:op}. In our specific case, this expression can be rewritten as~\cite{stammer_weak_2025}
\begin{equation}\label{Eq:App:intensity}
	\langle I(\vb{r},t)\rangle
		= \int \dd^2 E_\alpha Q(E_\alpha) \langle \chi_q(E_\alpha)\vert \hat{\Pi}_{i,q}(\vb{r},t)\vert \chi_q(E_\alpha)\rangle.
\end{equation}

Using the relation $\hat{D}^\dagger(\alpha)\hat{\Pi}_{i,q}(\vb{r},t)\hat{D}(\alpha) = \hat{\Pi}_{i,q}(\vb{r},t) + E_{q}(\vb{r},t,\alpha)$, where $E_{q}(\vb{r},t,\alpha) = \mel{\chi_q}{\hat{\Pi}_{i,q}(\vb{r},t)}{\chi_q}$, Eq.~\eqref{Eq:App:intensity} can be recast as 
\begin{equation}
	\langle I(\vb{r},t)\rangle
		= 
		 \int \dd^2 E_\alpha Q(E_\alpha)
		 	\Big[
		 		E^2_{q}(\vb{r},t,E_\alpha)
		 		+ 2 E_{q}(\vb{r},t,E_\alpha)  \hat{\Pi}_{i,q}(\vb{r},t)
		 		+  \hat{\Pi}^2_{i,q}(\vb{r},t)
		 	\Big].
\end{equation}
Provided that $\vert\!\vert\hat{\Pi}_{i,q}(\vb{r},t)\vert\!\vert\propto10^{-8}$ a.u. $\ll E_{q}(\vb{r},t,\alpha)$  this expression can be further approximated by
\begin{equation}
	\langle I(\vb{r},t)\rangle
		\approx 
		\int \dd E_\alpha Q(E_\alpha)
				E^2_{q}(\vb{r},t,E_\alpha).
\end{equation}

Applying the Ehrenfest theorem to Eq.~\eqref{Eq:App:evolv:Pi:op}, it follows immediately that $E_{q}(\vb{r},t,\alpha)$ can be obtained from Eq.~\eqref{Eq:App:intensity} when neglecting the commutator involving $\hat{H}_{\text{others}}$.~Furthermore, by applying the paraxial approximation, $E_{q}(\vb{r},t,E_\alpha)$ can be computed using Eq.~\eqref{Eq:paraxial:cl:prop}.\\

\noindent\textbf{Phase mismatch $\Delta k$}\\ 
The phase mismatch is given by with $\Delta k=\Delta k_{\text{med}}+\Delta k_{\text{el}}+\Delta k_{\text{foc}}+\Delta k_{\text{dip}}$, with contributions from medium dispersion (med), free electrons (el), focusing conditions (foc) and the atomic dipole (dip), respectively~\cite{Huillier_NatRevPhys_2022}.~In this work, we neglect the focusing and dipole terms, i.e., $\Delta k_{\text{foc}}\approx0$ and $\Delta k_{\text{at}}\approx0$, which is a good approximation for $d_{f}\gg L_{m}$ and for frequencies near a single harmonic order.~Thus, we consider $\Delta k\approx\Delta k_{\text{med}}+\Delta k_{\text{el}}$, with $\Delta k_{\text{med}}=q k_{0}-k_{q}$ and
\begin{equation}\label{Eq:PhaseElectron}
	\begin{split}
		\Delta k_{\text{el}}\approx -\frac{e^2\omega_{q} \rho_{at} Y_{i}(I)}{2 \epsilon_{0} c  m_{e} \omega^{2}}.
	\end{split}
\end{equation}
Here, $\sigma^{(1)}$, is the single-photon absorption cross section of Ar atoms induced by the harmonics, which is $\sim 10^{-17}$cm$^{2}$ in the 20-40 eV photon energy range \cite{Hanke_IonData_1993}. The wave numbers of the driving IR and the $q$th harmonic order are $k_{0}=n_{0} \omega_{0}/c$ and $k_{q}=n_{q} \omega_{q}/c$ respectively, with $n_{0}$ and $n_{q}$ the corresponding refractive indices. The electron mass is denoted as $m_{e}$. For the 15th harmonic, $\Delta k_{\text{med}}\approx 2 \times 10^{-6} (\text{rad/cm})$ \cite{Hanke_IonData_1993, Huillier_NatRevPhys_2022}.\\

\noindent\textbf{On the derivation of Eq.~\ref{Eq:HarmYield} }\\
In Eq.~\ref{Eq:HarmYield}, we consider that for plateau harmonics $(\langle N_{q} \rangle^{(\text{Coh.})})(\langle N_{q} \rangle^{(\text{BSV})})=(\langle N_{15} \rangle^{(\text{Coh.})})(\langle N_{15} \rangle^{(\text{BSV})})$. We also consider that $\langle N_{q} \rangle^{(\text{Coh.})}$ and $\langle N_{q} \rangle^{(\text{BSV})}$ are the harmonic photon numbers generated by the coherent and BSV states after propagation of $L_{m}\gtrsim\frac{5}{2}  L_{a}$ and $L_{m}\lesssim L_{m}^{(\text{BSV})}$, respectively. The power spectrum generated by coherent and BSV states after propagation of $L_{m}\approx\tfrac{5}{2}  L_{a}$ and $L_{m}\approx L_{m}^{(\text{BSV})}\approx2L_{a}$, respectively, is shown in Fig.~\ref{Fig.4}~(e).\\

\noindent\textbf{Photon number estimation of high harmonics generated by BSV}\\ 
To estimate the photon number of the high harmonics generated by BSV light, we use the experimentally and theoretically demonstrated values of conversion efficiency (defined as $\text{CE}_{S}=N_{q}/N_{\text{IR}}\propto S$) of the HHG process in Ar atoms ($S$ is the focal spot area of the laser beam in the HHG medium). For interactions of Ar atoms with $\approx 60$ fs IR laser pulses of $N^{(\text{coh.})}_{\text{IR}}\sim 2 \times 10^{16}$ photons per pulse,  $I \sim 10^{14} \text{W/}\text{cm}^{2}$, $\rho_{at} \sim 10^{18}$ atoms/cm$^3$ and $S_{\text{coh.}}\sim 5 \times 10^{-4}$ cm$^2$, it has been found \cite{Hergott_PRA_2002} that the $\text{CE}_{S_{\text{coh.}}}\sim 5 \times 10^{-6}$. Considering that the available BSV sources deliver pulses with mean photon number $\sim 10^{13}$ photons \cite{Spasibko_PRL-BSV_2017, Manceau_PRL-BSV_2019}, we have to focus a 60 fs BSV pulse into focal spot area of $S_{\text{BSV}}\approx 2.5 \times 10^{-7}$ cm$^2$ in order to reach an intensity of $\sim 10^{14} \text{W/}\text{cm}^{2}$. In this case the $\text{CE}_{S_{\text{BSV}}}\sim 2.5 \times 10^{-9}$ and the estimated harmonic photon number per pulse will be $\sim 2.5 \times  10^{4} \times \langle N_{15} \rangle^{(\text{BSV})}/\langle N_{15} \rangle^{(\text{Coh.})}$. This, for a $L_{m}\approx L_{m}^{(\text{BSV})}\approx2L_{a}$ long medium length, results that $\sim  5 \times 10^{2}$ harmonic photons per pulse exiting the Ar medium.

For BSV of mean photon number $\sim 10^{13}$ photons and mean intensity $\sim 10^{13} \text{W/}\text{cm}^{2}$, the photon losses due to ionization can be considered negligible i.e., $A\ll12\%$, the focal spot size can be increased to $S_{\text{BSV}}\approx 10^{-6}$ cm$^2$. Considering that for this BSV intensity, the $\langle N_{15} \rangle^{(\text{BSV})}$ is an order of magnitude lower than the corresponding value at $I_{sat}$, we obtain that $\text{CE}_{S_{\text{BSV}}}\sim  10^{-9}$ and $\langle N_{15} \rangle^{(\text{BSV})}/\langle N_{15} \rangle^{(\text{Coh.})}\sim 2 \times 10^{-3}$. Therefor the estimated photon number exiting the medium is $\sim  10^{4} \times \langle N_{15} \rangle^{(\text{BSV})}/\langle N_{15} \rangle^{(\text{Coh.})}=20$ photons per pulse.

\subsection{BSV photon losses due to scattering and HHG}\label{Sec:App:BSV:photon:losses}
The photon losses induced by IR scattering are on the order of $10^{-8}$ photons/cm and can be considered negligible. For $\varrho_{\text{at}}\sim 10^{18}$ atoms/cm$^3$, the photon losses induced by IR scattering are on the order of $10^{-8}$ photons/cm and can thus be considered negligible. Comparing the ionization probabilities (Fig.~\ref{Fig.2}(b)) with the harmonic generation probabilities (Fig.~\ref{Fig.4}(a)), we see that the major part of the IR photon losses is associated with the atomic ionization.

\subsection{Second-order autocorrelation and optical quadratures variance for a driver with photon losses}\label{Sec:App:g2:and:quadratures}
Generally, the initial state can be expressed in terms of the generalized positive $P$-representation as
\begin{equation}
	\hat{\rho}
		= \int \dd^2 \alpha \int \dd^2 \beta 
			\dfrac{P(\alpha,\beta^*)}{\braket{\beta^*}{\alpha}}
				\dyad{\alpha}{\beta^*}.
\end{equation}
Introducing photon losses modeled through a noisy beam splitter---an appropriate description for gaussian environments---we obtain for the reduced state of the driver
\begin{equation}
	\hat{\tilde{\rho}}
		= \int \dd^2 \alpha \int \dd^2 \beta
					\dfrac{P(\alpha,\beta^*)}{\braket{\beta^*}{\alpha}}
						\braket{\sqrt{1-t}\beta^*}{\sqrt{1-t}\alpha}
							\lvert\alpha\sqrt{t}\rangle\!\langle\beta^*\sqrt{t}\vert.
\end{equation}
From this expression it follows directly that any $n$-order autocorrelator can be written as
\begin{equation}
	\langle \hat{a}^{\dagger n}\hat{a}^n \rangle_{\text{noisy}}
		= t^{n}
			 \int \dd^2 \alpha \int \dd^2 \beta 
				\dfrac{P(\alpha,\beta^*)}{\braket{\beta^*}{\alpha}}
					\braket{\sqrt{1-t}\beta^*}{\sqrt{1-t}\alpha}
					\langle\beta^*\sqrt{t}\vert\alpha\sqrt{t}\rangle
					\beta^n \alpha^n.
\end{equation}

Interestingly, provided that
\begin{equation}
	\langle\beta^*\sqrt{t}\vert\alpha\sqrt{t}\rangle
		= \exp[-\dfrac{t}{2}
				\big(
					\abs{\alpha}^2 + \abs{\beta}^2
					- 2 \beta\alpha
				\big)],
\end{equation}
we immediately find that $\braket{\sqrt{1-t}\beta^*}{\sqrt{1-t}\alpha}\braket{\sqrt{t}\beta^*}{\sqrt{t}\alpha} = \braket{\beta^*}{\alpha}$. Hence, we obtain
\begin{equation}
	\langle \hat{a}^{\dagger n}\hat{a}^n \rangle_{\text{noisy}}
		= 
		 t^{n}
			\int \dd^2 \alpha \int \dd^2 \beta 
				P(\alpha,\beta^*)
					\beta^n \alpha^n
		= t^{n} \langle \hat{a}^{\dagger n}\hat{a}^n \rangle_{0}.
\end{equation}
and in particular the following expression for the $g^{(2)}(0)$ function
\begin{equation}
	g_{\text{noisy}}^{(2)}(0)
		= g_{0}^{(2)}(0).
\end{equation}

It is worth noting, however, that $g^{(2)}(0)$ becomes ill-defined when $t = 0$. In this case, both the numerator ($\langle \hat{a}^{\dagger 2} \hat{a}^2\rangle$) and denominator ($\langle \hat{a}^{\dagger} \hat{a}\rangle$) vanish, preventing a meaningful definition of $g^{(2)}(0)$. For intermediate values of $t$, although $g^{(2)}(0)$ may remain identical to that of the original state, this does not imply that the final state exhibits squeezing. To assess squeezing, it is necessary to evaluate the variances of the optical quadratures. Using
\begin{equation}
	\langle \hat{a} \rangle_{\text{noisy}}
		 = \sqrt{t}  \langle \hat{a} \rangle_{0},
\end{equation}
and defining $\hat{X}_1 = (\hat{a}^\dagger + \hat{a})$ and $\hat{X}_2 = -i (\hat{a}^\dagger - \hat{a})$, we obtain
\begin{equation}
	\langle \hat{X}_i\rangle_{\text{noisy}}
		= \sqrt{t} \langle \hat{X}_i\rangle_0,
	\quad
	\langle \hat{X}^2_i\rangle_{\text{noisy}}
		= t \big(
				\langle \hat{X}^2_i\rangle_0 -1
			\big) + 1,
\end{equation}
so that the variances read
\begin{equation}
	\Delta X_i^2\vert_{\text{noisy}}
		= t \Delta X_i^2\vert_0 + (1-t),
\end{equation}
with $\Delta X_1^2\vert_0 = e^{2r}$ and $\Delta X_2^2\vert_0 = e^{-2r}$ for a BSV state with squeezing parameter $r$.

\begin{figure}
    \centering
    \includegraphics[width = 1\textwidth]{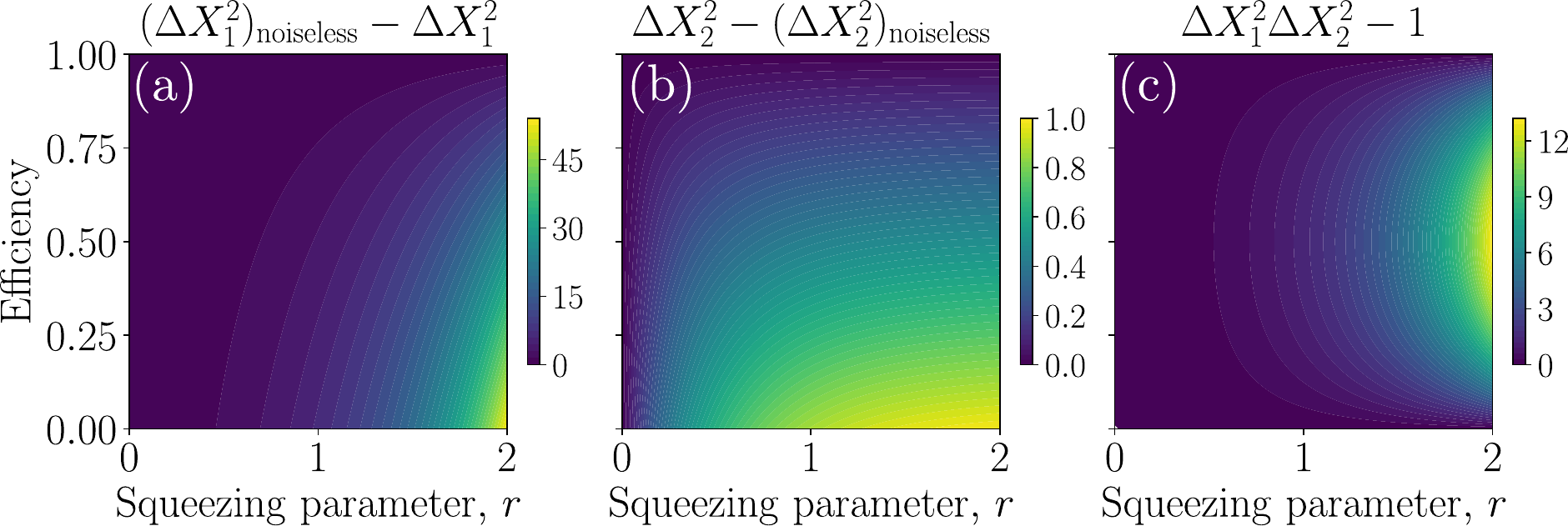}
    \caption{(a) Variance of $\hat{X}_1$ for the noiseless state compared to the same state with photon losses. (b) Same as (a), but for $\hat{X}_2$ and multiplied by a minus sign. (c) Product of both variances minus the Heisenberg limit.}
    \label{fig:BSV:props}
\end{figure}

Figure~\ref{fig:BSV:props} shows the variances along different optical quadratures (panels (a) and (b)), as well as the products of these variances (panel (c)).~Several features can be highlighted:
\begin{itemize}
    \item \textbf{Panel (a).}~Displays the variance of $X_1$ (antisqueezed quadrature), computed as the difference between that of the noiseless state and the noisy one.~As $t$ increases, the distribution becomes more concentrated compared to the ideal, lossless case.
    \item \textbf{Panel (b).}~Shows the variance of $X_2$ (squeezed quadrature) of the noisy state minus that of the noiseless one---the opposite of panel (a).~This difference is always positive, indicating that squeezing is gradually lost along this quadrature as photon losses increase.
    \item \textbf{Panel (c).}~Displays the product of both variances minus 1, corresponding to the Heisenberg limit in our notation.~We observe the product of the variances exceeds one for noisy states and reaches zero only at the extremes (either a pure BSV or a vacuum state).~In between, the state is noisy:~although it retains some stretching reminiscent of squeezing, these features do not correspond to genuine \emph{quantum squeezing}.
\end{itemize}

\section{On the calculation of $\langle N_{q} \rangle$}
\begin{figure*} [h!]
	\centering
	\includegraphics[width=0.5\textwidth]{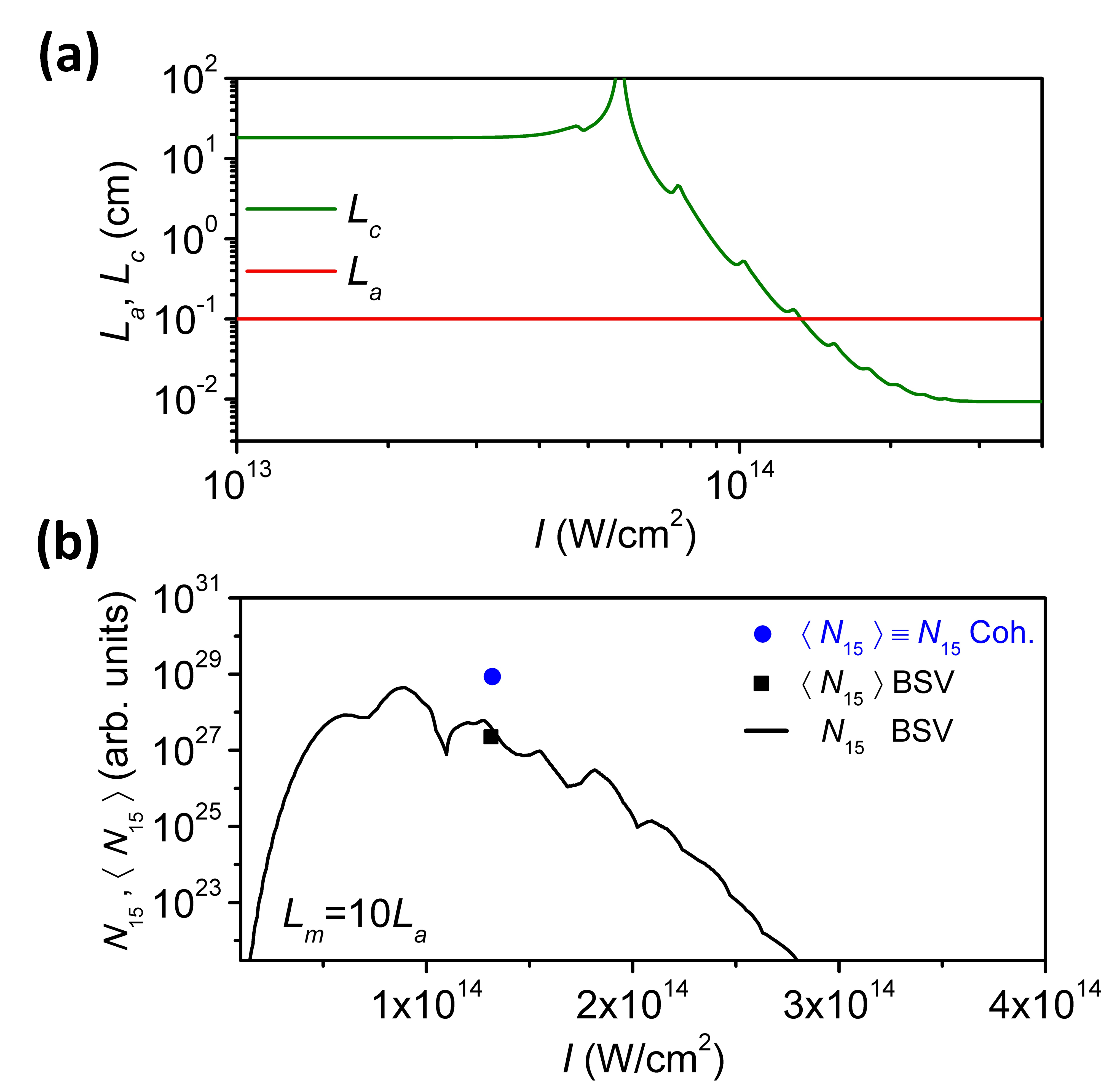}
	
    \caption{(a) The green-line shows the dependence of the coherence length $L_{c}$ on the intensity of driving field $I$. The red-line is the constant absorption length $L_{a}$. It is noted that, $I$ corresponds to each intensity value in the $Q(E)$ distribution of a BSV light with $\langle I \rangle=I_{\text{sat}}$ ($x$-axis in Fig.2(a) of the main text of the manuscript). (b) The black-line shows the dependence of $N_{15}$ on each intensity $I$ in the $Q(E)$ distribution of the BSV field, for $L_{m}=10 L_{a}$. The black and blue points show the $\langle N_{15} \rangle$ of the BSV and coherent states, respectively.}
	\label{Fig.S3}
\end{figure*}

In Fig.~\ref{Fig.S3}, we illustrate the calculations using the 15th harmonic, and a gas density of $\rho_{\text{at}} \sim 10^{18}$ atoms/cm$^3$, typical for HHG experiments. The mean intensity was set at $\langle I\rangle =I_{\text{sat}}$. Figure~\ref{Fig.S3}(a) shows the dependence of $L_{c}$ (green curve) and $L_a$ (red line) on the driving intensity. Here, $I$ corresponds to the intensity values sampled from the $Q(E)$ distribution of a BSV field with $\langle I \rangle = I_{\text{sat}}$. The black line in Fig.~\ref{Fig.S3}(b) displays the dependence of $N_{15}$ on $I$ across the $Q(E)$ distribution for a BSV field, computed for $L_{m}=10 L_{a}$. The maximum $N_{15}$ occurs around $ I \approx 9 \times 10^{13} \text{W}/\text{cm}^{2}$. At higher intensities, $N_{15}$ drops sharply due to strong ionization. Averaging over $Q(E)$ yields $\langle N_{15} \rangle$ (black point), which is about thirty times smaller than the value obtained for a coherent state (blue point). Finally, Fig.~\ref{Fig.4}(c) of the main text shows $\langle N_{15} \rangle$ as a function of the medium length $L_{m}$ for BSV (black-line) and coherent (blue-line) state driving fields.

\end{document}